\documentclass[aps,pre,reprint,groupedaddress,showpacs,preprintnumbers,amsmath,amssymb,twocolumn]{revtex4-1}

\usepackage{color}
\usepackage{dcolumn}
\usepackage{bm}
\usepackage{hyperref}
\usepackage{mathptmx}
\usepackage{graphicx,epsfig,epsf}
\usepackage{subfigure}
\usepackage{caption}

\usepackage{color}
\begin{document}
\title{Effect of cohesion on shear banding in quasi-static granular material}
\author{Abhinendra Singh}
\email[]{a.singh-1@utwente.nl}
\affiliation{Multi Scale Mechanics (MSM), Faculty of Engineering Technology, MESA+, University of Twente, P.O. Box 217, 7500 AE Enschede, The Netherlands.}
\author{Vanessa Magnanimo}
\affiliation{Multi Scale Mechanics (MSM), Faculty of Engineering Technology, MESA+, University of Twente, P.O. Box 217, 7500 AE Enschede, The Netherlands.}
\author{Kuniyasu Saitoh}
\affiliation{Multi Scale Mechanics (MSM), Faculty of Engineering Technology, MESA+, University of Twente, P.O. Box 217, 7500 AE Enschede, The Netherlands.}
\author{Stefan Luding}
\affiliation{Multi Scale Mechanics (MSM), Faculty of Engineering Technology, MESA+, University of Twente, P.O. Box 217, 7500 AE Enschede, The Netherlands.}
\date{\today}
\begin{abstract}
It is widely recognized in particle technology that adhesive powders show a wide range of different bulk behavior due to the peculiarity of the particle interaction.
We use Discrete Element simulations to investigate the effect of contact cohesion on the steady state of dense powders in a slowly sheared split-bottom Couette cell,
which imposes a wide stable shear band. The intensity of cohesive forces can be quantified by the {\em granular Bond number} ($Bo$), namely the ratio
 between maximum attractive force and average force due to external compression.
We find that the shear banding phenomenon is almost independent of cohesion for Bond numbers $Bo<1$,
but for $Bo \ge 1$ cohesive forces start to play an important role, as both width and center position of the band increase for $Bo > 1$.
Inside the shear band, the mean normal contact force is always independent of cohesion and depends only on the confining stress.
In contrast, when the behavior is analyzed focusing on the eigen-directions of the local strain rate tensor, a dependence on cohesion shows up.
Forces carried by contacts along the compressive and tensile directions are symmetric about the mean force (larger and smaller respectively),
while the force along the third, neutral direction follows the mean force. This anisotropy of the force network increases with cohesion, just like the heterogeneity in all
(compressive, tensile and neutral) directions.

\end{abstract}
\maketitle
\section{Introduction}
Granular materials such as sand and limestone, neither behave like elastic solids nor like normal fluids, which makes their motion difficult to predict.
 When they yield under slow shear, the relative motion is confined to narrow regions (between large solid-like parts) called shear bands \cite{bridgwater80,howell99,schall2010shear}.
 Shear bands are observed in many complex materials, which range from foams \cite{Katgert08} and emulsions \cite{Addad05,Ovarlez09}
 to colloids \cite{Dhont03} and granular matter \cite{Mandl77,bridgwater80,bardet91,oda98,Scott96,muhlhaus87b,latzel00,latzel03,fenistein03,mcnamara1996dynamics,mcnamara1994inelastic,howell99}.
 There has been tremendous effort to understand the shear banding in flow of non-cohesive
 grains \cite{Mandl77,bridgwater80,bardet91,oda98,Scott96,muhlhaus87b,latzel00, latzel03,fenistein03,mcnamara1996dynamics,mcnamara1994inelastic,howell99,KamrinKoval2012,kamrin2013predictive}.
 However, real granular materials often experience inter-particle attractive forces due to many physical phenomena: {\em van der} Waals due to atomic
 forces for small grains \cite{Quintanilla03, Valverdejammingpowder04, Castellanos05}, {\em capillary} forces due to presence of humidity \cite{herminghaus05}, {\em solid bridges}
 \cite{Micela05,brendel-2011}, coagulation of particles \cite{hatzes91}, and many more. 

The question, arises how does the presence of attractive forces affect shear banding?
So far, only a few attempts have been made to answer this question, concerning dense metallic glasses \cite{Spaepen1977,LiSBglasses}, adhesive emulsions \cite{Becu06,Chaudhuri2012},
 attractive colloids \cite{vermant2001large,hohler2005rheology,coussot2010physical}, cemented granular media \cite{Nicolas10}, wet granular media \cite{Roman12,Fabian13} and clayey soils \cite{Yuan13}.
 Recently, rheological studies on adhesive emulsions and colloids
 \cite{Becu06,Chaudhuri2012,vermant2001large,coussot2010physical} reported that the presence of attractive forces 
 at contact affects shear banding by affecting flow heterogeneity and wall slip.

Another unique yet not completely understood feature of granular materials is their highly heterogeneous contact
 force distribution. The heterogeneity in the force distribution has been observed in both experimental and numerical studies
 \cite{liu1995force,mueth98,howell99,lovoll0000force,blair01b,majmudar2005contact,liffman1992force,radjai98b,silbert2002statistics,Snoeijer03}.
While huge effort has been made to understand the force distribution of non-cohesive particles 
\cite{liu1995force,mueth98,howell99,lovoll0000force,blair01b,majmudar2005contact,liffman1992force,radjai98b,Silber10},
 only limited studies have aimed to understand the same for assemblies with attractive interactions 
\cite{trappe2001jamming,Valverdejammingpowder04,Radjaiwet06,Rouxcoh07,ABYucohfor08,Radjaidry-wet10}. 
Richefeu et al.\ \cite{Radjaiwet06} studied the stress
 transmission in wet granular system subjected to isotropic compression. Gilabert et al.\ \cite{Rouxcoh07} focussed on a two-dimensional packing
 made of particles with short-range interactions (cohesive powders) under weak compaction.
Yang et al.\ \cite{ABYucohfor08} studied the effect of cohesion on force structures in a static granular packing by changing the particle size.
In a previous study \cite{luding11}, the effect of dry cohesion at contact on the critical state yield stress was studied. The critical-state
 yield stress shows a peculiar non-linear dependence on the confining pressure related to cohesion.
 But the microscopic origin was not studied.

 In this paper, we report the effect of varying attractive forces at contact on the steady state flow
 behavior and the force structure in sheared dry cohesive powders.
 Discrete Element Method (DEM) simulations are used to investigate the system at micro (partial) and macro level.
 In order to quantify the intensity of cohesion, a variation of the {\em granular Bond number} \cite{Nase01,Rouxcoh07,Rognon08} is introduced.
 We find that this dimensionless number very well captures the transition from a gravity/shear-dominated regime to the cohesion-dominated regime.
 To understand this further we look at the effect of cohesion on the mean force and anisotropy, by investigating the forces along the eigen-directions
 of the local strain rate tensor.
 Intuitively, one would expect only the tensile direction to be affected by cohesion, but the real behavior is more complex.
 We also discuss the probability distributions and heterogeneities of the forces in different directions to complete the picture.

The paper is organized in four main parts. Section \ref{sec:modelgeo} describes the model system 
in detail specifying the geometry, details of particle properties, and the interaction laws.
In section \ref{sec:results}, the velocity profiles and shear band from samples with different contact cohesion
 are presented. In the same section, the force anisotropy and probabilities are studied too.
Finally, section \ref{sec:dis} is dedicated to the discussion of the results, conclusions and an outlook.
\section{Discrete element method simulation (DEM)}\label{sec:modelgeo}
In this section, we explain our DEM simulations.
We introduce a model of cohesive grains in Sec.\ \ref{sub:model} and show our numerical setup in Sec.\ \ref{sub:setup}.
In Sec.\ \ref{sub:param}, we introduce a control parameter, 
i.e.,\ the \emph{global Bond number}, which governs the flow profiles and 
structure of the system.
\subsection{Model}\label{sub:model}

DEM provides numerical solutions of Newton's equations of motion based on the specification of particle properties viz.\ stiffness,
density, radius and a certain type of interaction laws like Hertzian/Hookean \cite{allen87,cundall71}.
Simulation methodology and material parameters used in this study are the same as in our previous work \cite{luding11,asingh13}. 
The adhesive elasto-plastic contact model \cite{Luding08gm} is used to simulate cohesive bulk flow, 
as briefly explained below.

For fine, dry powders, adhesive properties due to van der Waals forces and plasticity and irreversible deformation in the vicinity of the contact
 have to be considered at the same time \cite{tomas2004fundamentals,thornton1998theoretical}.
 This complex behavior is modeled using a piece-wise linear hysteretic spring model \cite{Luding08gm}. Few other contact models
 in similar spirit are also recently proposed \cite{thakur2013experimental,pasha2014linear}.

The adhesive, plastic (hysteretic) force is introduced by allowing the normal unloading stiffness to depend on the history of deformation.
During initial loading the force increases linearly with overlap $\delta$ along $k_1$, until the maximum overlap $\delta_{\rm max}$
 is reached, which acts as a history parameter. During unloading the force decreases along $k_2$, the value of which depends on the maximum overlap $\delta_{\rm max}$
 as given by Eq.\ \eqref{eq:k2_delmax}. The overlap when the unloading force reaches zero, $\delta_0=(1-k_1/k_2)\delta_{\rm max}$,
 resembles the permanent plastic deformation and depends nonlinearly on the previous maximal force $f_{\rm max}=k_1\delta_{\rm max}$.
 The negative forces reached by further unloading are attractive, cohesion forces, which also increase nonlinearly with the previous
 maximum force experienced. The maximal cohesion force that corresponds to the \textquotedblleft pull--off\textquotedblright\ force,
is given by
\begin{equation}
 f_{\rm m}=-k_c \delta_{\rm min}, 
\label{eq:fmin_del}
\end{equation}
 with 
$\delta_{\rm min}=\frac{k_2-k_1}{k_2+k_c}\delta_{\rm max}$.
\begin{figure}
\centering
\includegraphics[width=6cm]{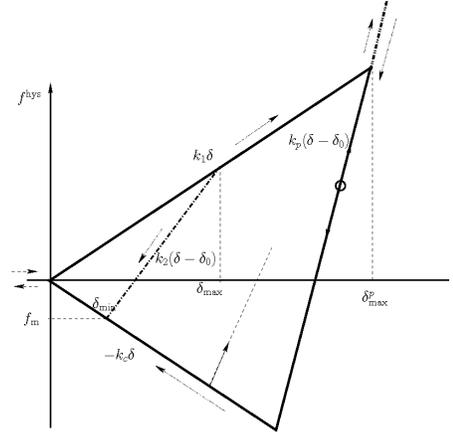}
\caption{Schematic graph of the piece-wise linear, hysteretic, 
         and adhesive force-displacement model in normal direction.
}
\label{fig:fnonadh}
\end{figure}

Three physical phenomena: elasticity, plasticity and cohesion are quantified
 by three material parameters $k_p$, $k_1$, and $k_c$, respectively.
 Plasticity disappears for $k_1=k_p$ and cohesion vanishes for $k_c=0$.
%
%
In the following we focus on the relative importance of cohesion
and thus do not provide measurable force magnitudes.
Furthermore, the contact model has to be seen as a meso-scale model, 
where each particle represents an ensemble of primary particles 
and the contact model represents the respective bulk behavior, 
see Ref.\ \cite{asingh14}, without a direct match of
the magnitude of forces in the model with the forces between the 
primary particles. Qualitatively, the interpretation of $k_c$ 
is that it describes the increased van der Waals type adhesion due
to plastic deformations (both of the particles and the micro-structure)
under compression, which increase the contact surface and thus the cohesion.
Some considerations on the magnitude and relative importance of the
cohesion force can be found in Appendix \ref{sec:fmin_magnitude}.
%
%

In order to account for realistic load-dependent contact behavior, the $k_2$ 
value is chosen to depend on the maximum overlap  $\delta_{\rm max}$, 
i.e.\ {\em particles are more stiff for larger previous deformation} and so the dissipation
is dependent on deformation.  The dependence of $k_2$ on overlap $\delta_{\rm max}$ is chosen 
empirically as linear interpolation
\begin{equation}
k_2(\delta_{\rm max}) = \left \{
\begin{array}{lll}
k_{p}\,{\rm~~~~~~~~~~~~~~~~~~~~~~~~if~~} \delta_{\rm max} / \delta^{p}_{\rm max} \ge 1 \\
k_1 + (k_{p}-k_1)\frac{{\delta_{\rm max} }}{{ \delta^{p}_{\rm max} }}    & \\
~~~~~~~~~~{\rm~~~~~~~~~~~~~~~~~~~if~~} \delta_{\rm max} / \delta^{p}_{\rm max} < 1
\end{array}
\label{eq:k2_delmax}
\right . 
\end{equation}
 As discussed in Ref.\ \cite{Luding08gm}, very large deformations will
 lead to a quantitatively different contact behavior, a maximal force
 overlap $\delta^p_{\rm max} = \frac{k_{p}}{k_{p}-k_1}  \frac{2{a_1 a_2}}{a_1+a_2} \phi_f$
 is defined (with $\phi_f=0.05$). Above this overlap $k_2$ does not increase
 anymore and is set to the maximal value $k_2=k_p$. This visco-elastic,
 reversible branch is referred to as \textquotedblleft limit branch\textquotedblright.


The contact friction is set to $\mu=0.01$, i.e. artificially small, in order to be able to focus on the effect of contact cohesion only.
In order to study the influence of contact cohesion, we analyzed the system for the following set of adhesivity parameters $k_c$:
\begin{equation}
 k_c \in \left [ 0, 5, 10, 25, 33, 50, 75, 100, 200 \right ] \mathrm {Nm^{-1}}~,
\label{eq:k_c}
\end{equation}
which has to be seen in relation to $k_1=100$\  $\mathrm {Nm^{-1}}$.
Other parameters, such as the jump--in force $f_a=0$ \cite{asingh14} 
and $\phi_f=0.05$ \cite{asingh14} are not varied here.
 We also introduce damping forces proportional to the normal and tangential
 relative velocities, where the viscous coefficients are given by $\gamma_n = 0.002$\,s$^{-1}$ 
and $ \gamma_t =0.0005 s^{-1}$ respectively.
%
\subsection{Split-bottom ring shear cell}\label{sub:setup}
%
\begin{figure}

\includegraphics[width=6cm]{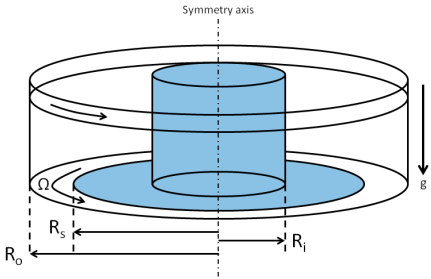}
\centering
\caption{(Color online)
A sketch of our numerical setup consisting of a fixed inner part (light blue shade) and a rotating outer part (white).
The white part of the base and the outer cylinder rotate with the same angular velocity $\Omega$ around the symmetry axis.
The inner, split, and outer radii are given by $R_i=0.0147$\,m, $R_s=0.085$\,m, and $R_o=0.11$\,m, respectively, where each radius is measured from the symmetry axis.
The gravity $g$ points downwards as shown by arrow.}
\label{fig:sketch_sp}
\end{figure}
%
Figure \ref{fig:sketch_sp} is a sketch of our numerical setup
 (as introduced in Refs.\ \cite{fenistein03,fenistein04,luding08p,luding08crs,dijksman10}).
In this figure, the inner, split, and outer radii are given by $R_i$, $R_s$, and $R_o$, respectively,
where the concentric cylinders rotate relative to each other around the symmetry axis (the dot-dashed line).
The ring shaped split at the bottom separates the moving and static parts of the system,
where a part of the bottom and the outer cylinder rotate at the same rate.
The system is filled with $N\approx 3.7\times10^{4}$ spherical particles with density $\rho = 2000$\,kg/m$^3 = 2$\,g/cm$^3$ up to height $H$.
The average size of particles is $a_0=1.1$\,mm, and the width of the homogeneous 
size-distribution (with $a_{\rm min}/a_{\rm max}=1/2$) is $1-{\cal A} = 1-\langle a \rangle^2/\langle a^2 \rangle = 0.18922$.
The cylindrical walls and the bottom are roughened due to some (about $3\%$ of the total number) attached/glued particles \cite{luding08p,luding08crs}.

When there is a relative motion at the split, a shear band propagates from the split position $R_s$ upwards
and inwards, remaining far away from cylinder-walls and bottom in most cases.
The qualitative behavior is governed by the ratio $H/R_s$ and three different regimes can be identified, 
as reported in Refs.\ \cite{Unger04shear, Fenistein06, ries07,dijksman10}.
We keep ${H}/{R_s} < 0.5$, such that the shear band reaches the free surface and 
stays away from inner wall \cite{Unger04shear,Fenistein06}.

Translational invariance is assumed in the azimuthal $\theta-$direction, and the averaging
 is performed over toroidal volumes, over many snapshots in time. This leads to
 fields $Q(r,z)$ as function of the radial and vertical positions.

Since we are interested in the quasi-static regime, the rotation rate of the outer cylinder 
is chosen to be $0.01$\,s$^{-1}$, such that the inertial number
$I=\frac{\dot{\gamma} d}{\sqrt{p/\rho}}$ \cite{da2005rheophysics} is $I\ll 1$, and the simulation runs for more than $50$\,s.
\subsection{Bond number}\label{sub:param}
Intensity of cohesion can be quantified by the ratio of the maximum attractive force to a typical force scale in the system.
For example, Nase et al.\ \cite{Nase01} introduced the granular Bond number under gravity,
which compares the maximum attractive force at contact with the weight of a single grain.
For plane shear without gravity, other authors \cite{Rouxcoh07,Rognon08} used a ratio between the maximum attractive force and the average force due to the confining pressure.
In our analysis, we introduce a \emph{global Bond number} as
\begin{equation}
Bo=\frac{f_{\rm m}}{\langle f \rangle }~, \label{eq:Bog}
\end{equation}
where $f_{\rm m}$ and $\langle f \rangle$ are the maximum allowed attractive force reached at a contact (given by the contact model, Appendix \ref{sec:fmin_press} using $\delta_{\rm max}= \delta^{p}_{\rm max}$ )
and the mean force per contact reached close to the bottom, respectively.
For the calculation of the mean force $\langle f \rangle$, a layer of two particle diameters away from the bottom
 is chosen. 
Because the shear band initiates from the bottom, we choose the mean force $\langle f \rangle$ close to the bottom
to understand the effect of cohesion on these shear bands.

It is important to mention that the mean compressive force (at the bottom) corresponds to the weight of the material above, whereas
 the maximum attractive force corresponds to the pull-off force,
 which is directly related to the surface energy of the particles. These two material and particle properties are easily accessible experimentally, see Appendix \ref{sec:fmin_magnitude}.

 The Bond number is a measure of the relative importance of adhesive forces compared to compressive forces.
  A low Bond number indicates that the system is relatively unaffected by attractive forces; a high number
 (typically larger than one) indicates that attractive forces dominate. Intermediate numbers indicate a non-trivial competition between the two effects.

In parallel with the global Bond number, we also define two local variants of this quantity.
 A local simulation based Bond number $Bo^s_{l}(p)=f^s_{\rm m}(p)/\langle f(p)\rangle$ can be defined by comparing the maximum attractive force reached at a given pressure
(which can be less than or equal to the maximum allowed attractive force given by the contact model) with the mean force at that pressure (subscript $l$ represents
the local quantity, while superscript $s$ denotes that this definition takes input from simulation data). Another variant of this 
$Bo^a_{l}(p)$ is defined in Appendix \ref{sec:fmin_press}, 
which compares the analytical prediction for the maximum attractive force with mean force at that pressure,
and do not use the gravitational Bond number, see Appendix \ref{sec:fmin_magnitude}, since it is only relevant close to the free 
surface and for single particles in contact with a wall.
%
%
%

\begin{figure}
\includegraphics[scale=0.35,angle=-90]{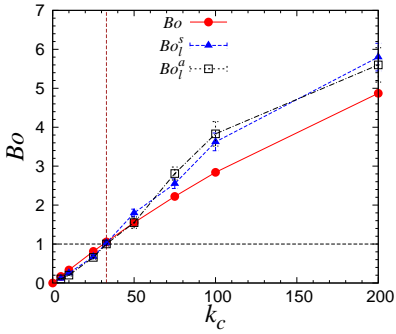}
\caption{(Color online)
Variants of granular Bond number plotted against cohesive strength $k_c$, where the red circles represent the global Bond number $Bo$,
while the blue triangles and black squares represent the average values of $Bo^s_{l}(p)$ and $Bo^a_{l}(p)$, respectively.}
\label{fig:Bond_number}
\end{figure}

Figure \ref{fig:Bond_number} displays the global Bond number $Bo$ and the mean values of $Bo^s_{l}(p)$ and $Bo^a_{l}(p)$
(averaged over different pressure) as functions of the adhesivity parameter $k_c$, where the figure shows that local
 and global quantities are comparable with slight divergence for high cohesion $k_c$.
For the sake of simplicity in the rest of this paper, we use the global Bond number $Bo$ to quantify the intensity of cohesion.

%
%
\section{Results}\label{sec:results}
%
In this section, we present our results of DEM simulations.
In Sec.\ \ref{sub:rheol}, we analyze the flow profiles and shear banding in the system.
In Sec.\ \ref{sub:struc}, we study distributions and structures of force chain networks in shear bands.
In Sec.\ \ref{sub:aniso}, we explain anisotropic features of the force chain networks.
%
\subsection{Effect of cohesion on Flow Profiles}
\label{sub:rheol}
%
\begin{figure}
\noindent
\includegraphics[scale=0.45]{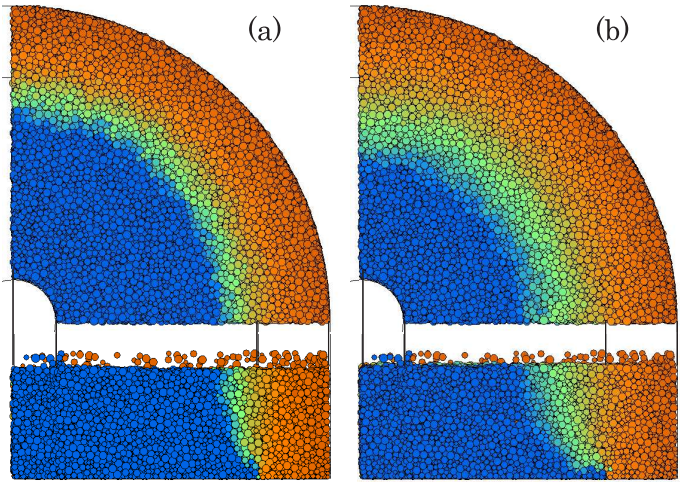}
\caption{\small
(Color online)
Snapshots from simulations with different cohesion strengths, but the same number of mobile particles $N=34518$,
 seen from the top (Top) and from the front (Bottom).
The material is (a) without cohesion $Bo=0$, and  (b) with strong cohesion $Bo=4.86$.
The colors blue, green, and orange denote the particles with displacements in tangential direction per second
$r \,{\rm d}\phi \le 0.5$\,mm, $r \,{\rm d}\phi \le 2$\,mm, $r \,{\rm d}\phi \le 4$\,mm, and $r \,{\rm d}\phi   > 4$\,mm, respectively}
\label{fig:shear11}
\end{figure}

Figure \ref{fig:shear11} displays both top- and front-view of samples with same filling height, i.e. same number of particles,
and different global Bond numbers, $Bo=$ (left) $0$ and (right) $4.86$, respectively,
where the color code represents the azimuthal displacement rate of the particles.
From the front-view, we observe that the shear band (green colored area) moves inwards and gets wider with increasing height and Bond number.

\begin{figure} 
\includegraphics[scale=0.35,angle=-90]{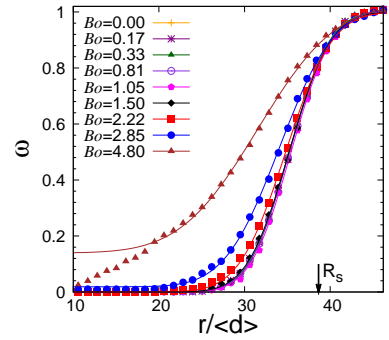}
\caption{(Color online)
Non-dimensional angular velocity profile $\omega$ at the top surface plotted against the radial coordinate $r$ scaled by the mean diameter $\langle d\rangle$.
Different symbols represent different values of the global Bond number $Bo$ given in the inset, where the solid lines represent the corresponding fits to Eq.\ \eqref{eq:v_phi}.} 
\label{fig:top_vel_coh}
\end{figure}

\begin{figure}
\subfigure[]{\includegraphics[scale=0.35,angle=-90]{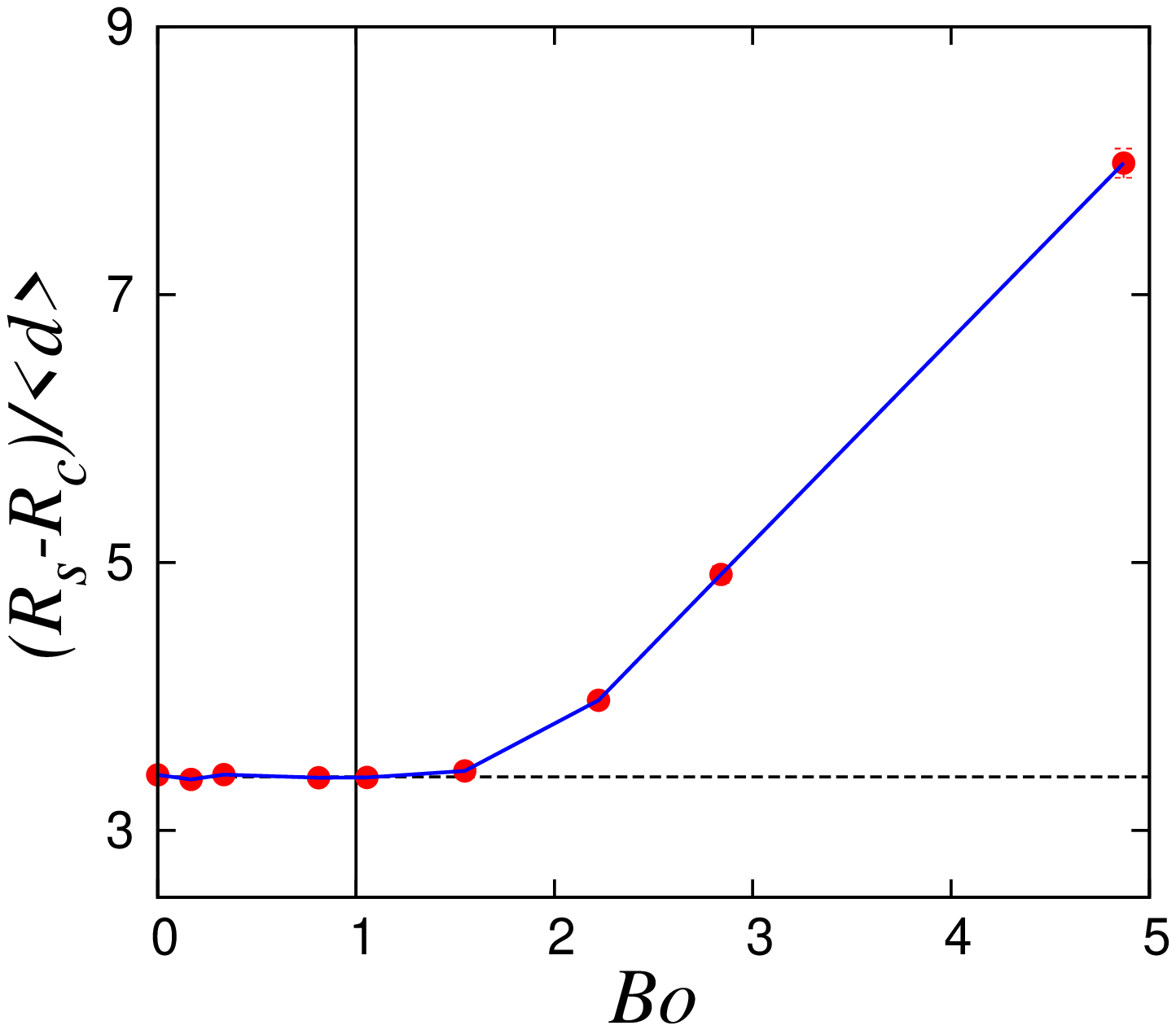}\label{fig:pos_top_coh}}
\subfigure[]{\includegraphics[scale=0.35,angle=-90]{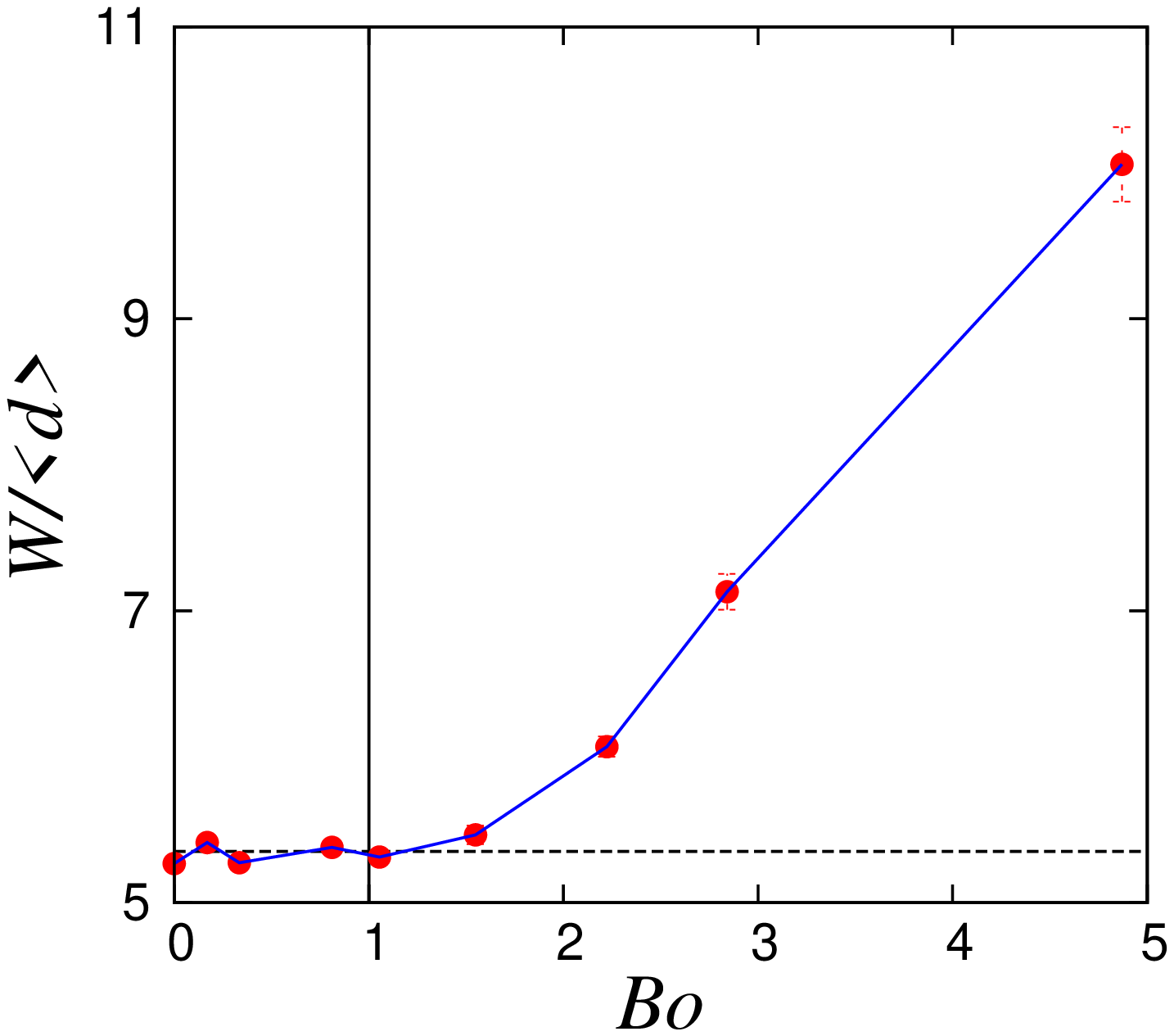}\label{fig:wid_top_coh}}
\caption{(Color online)
(a) Position and (b) width (both scaled by mean particle diameter) of shear band at the top surface plotted against the global Bond number $Bo$.
Symbols with error-bars are the data, while the lines are only a guide to eye.}
\label{fig:pw}
\end{figure}

With the goal to extract quantitative data for the shear band area, in Fig.\ \ref{fig:top_vel_coh} we plot the non-dimensional
 angular velocity profiles at the top surface against radial coordinate normalized with the mean particle diameter
 $\langle d \rangle$, where we assume translational invariance in the azimuthal direction and take averages over the toroidal volumes as well as many snapshots in time \cite{latzel00}.
The angular velocity profile can be well approximated by an error function
\begin{equation}
\omega= A_1 + A_2 {\rm erf}{  \left( \frac{r-R_c}{W} \right)}
\label{eq:v_phi}
\end{equation}
as in the case of non-cohesive materials \cite{luding08p,luding08crs,dijksman10,fenistein04,fenistein03},
where $R_c$ and $W$ are the position and width of the shear band, respectively.
Here, we use the dimensionless amplitudes, $A_1=A_2\approx 0.5$, for the whole range of the Bond numbers,
while we use $A_1=0.6$ and $A_2=0.4$ for the strong cohesion with $Bo=4.86$.
The dimensionless amplitudes, $A_1$ and $A_2$ (along with estimated errors), are summarized in Table \ref{tab:fitparameters}.
 Then we extract the position of the shear band relative to the split at bottom $R_s-R_c$
and the width of shear band $W$ (both scaled by mean particle diameter) at the top surface
 and we plot them in Fig.\ \ref{fig:pw} against the Bond number.
Within the error-bars, both the position and width are independent of cohesion if $Bo<1$.
However, the shear band moves inside and becomes wider with the Bond number for $Bo>1$.


\begin{figure}
\subfigure[]{\includegraphics[scale=0.35,angle=-90]{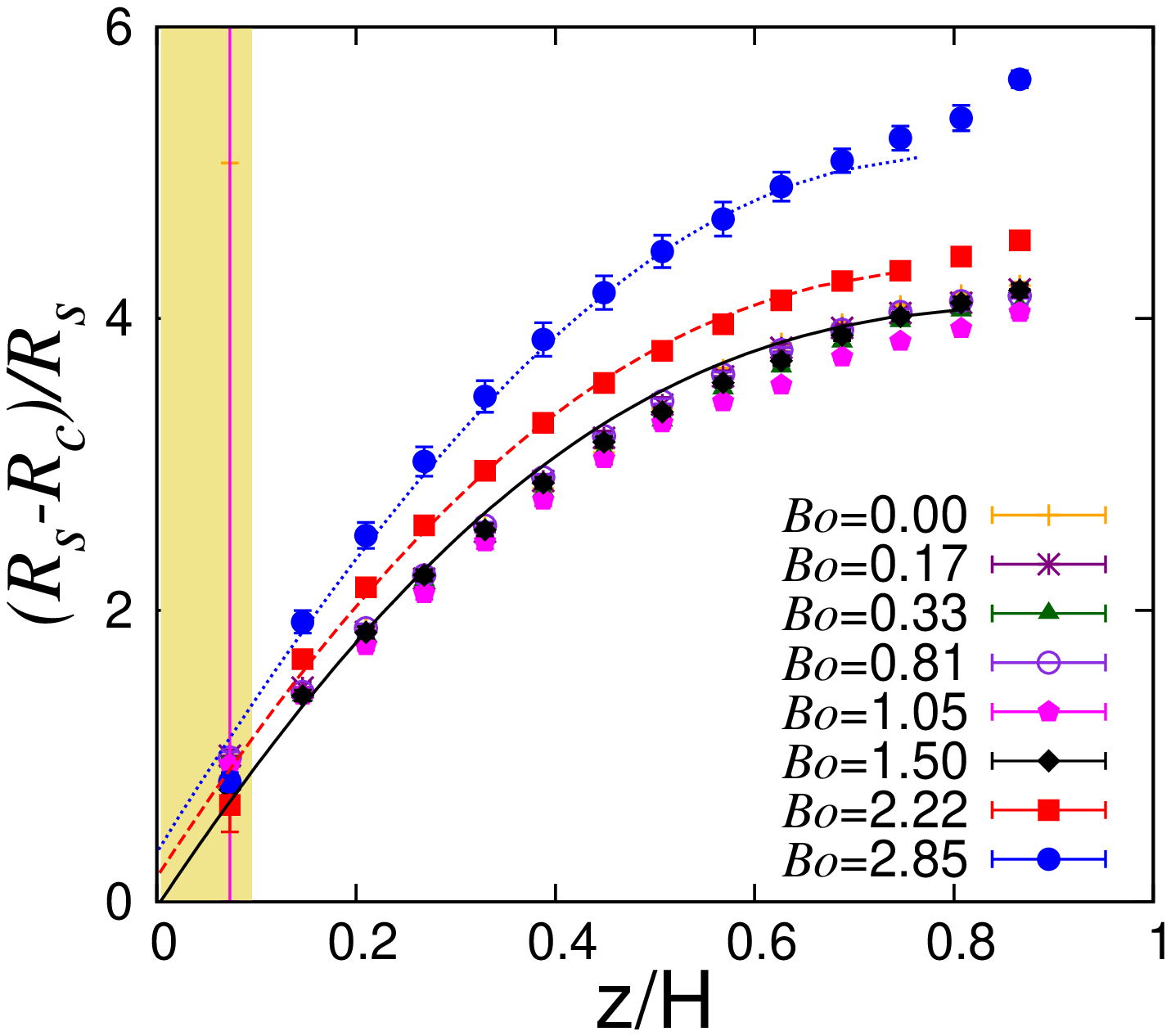}\label{fig:pos_coh}}\\
\subfigure[]{\includegraphics[scale=0.35,angle=-90]{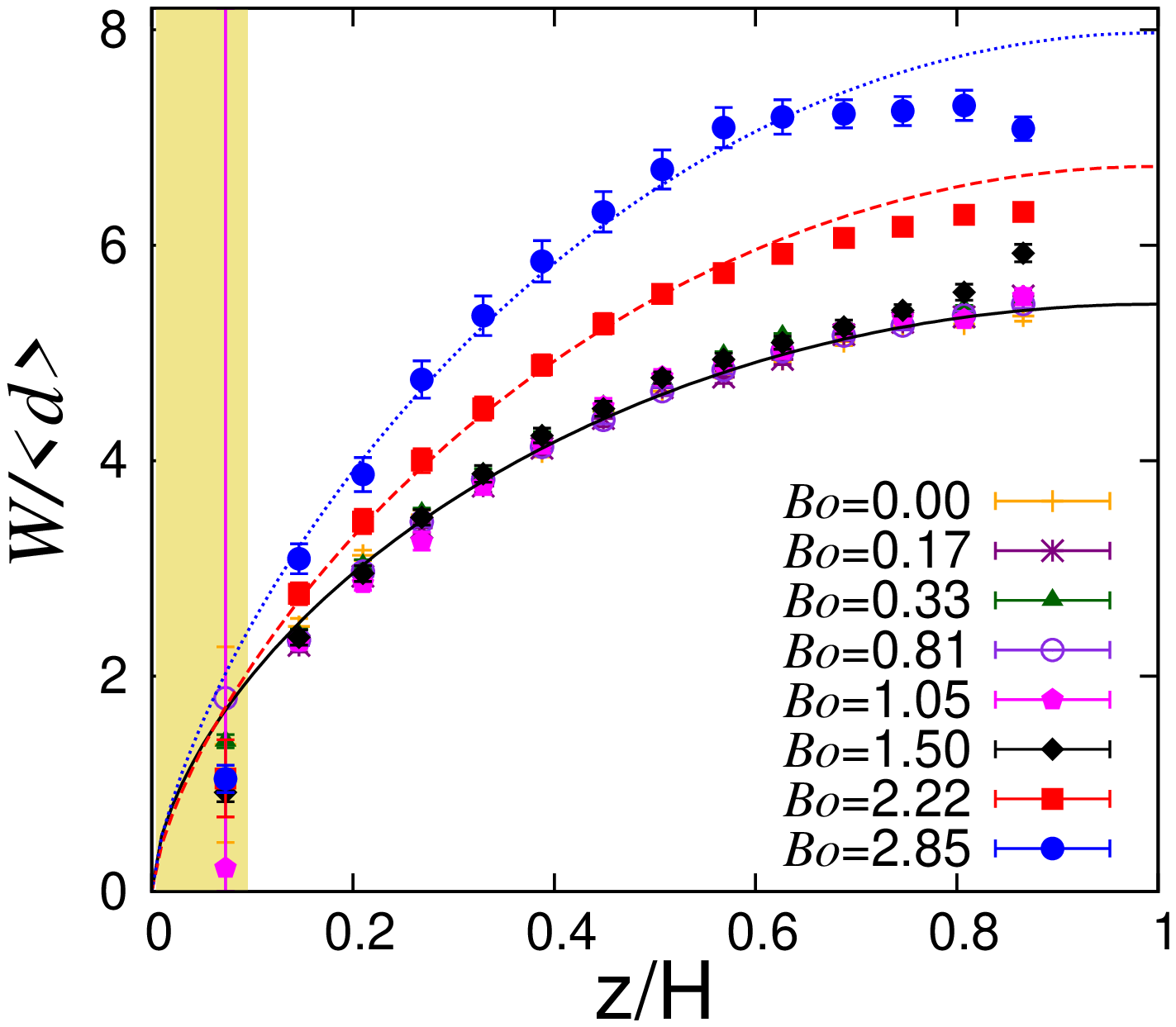}\label{fig:wid_coh}}
\caption{(Color online)
(a) Position and (b) width (both scaled by mean particle diameter) of shear band in the cell plotted against height $z$ scaled by the filling height $H$.
Different symbols correspond to values of the global Bond number $Bo$ given in the inset.
The lines in (a) and (b) are the predictions, Eqs.\ \eqref{eq:Rc-z} and \eqref{eq:W-z}, respectively.}
\label{fig:SB_coh_comp}
\end{figure}

Both $R_s-R_c$ and $W$ also depend on the height ($z$) in the system.
Figure \ref{fig:SB_coh_comp} displays the non-dimensional position and width of the shear band
for different values of $Bo$ as functions of the height scaled by the filling height, i.e. $z/H$.
The shear band moves closer to the inner cylinder and gets wider while approaching the top layer,
which is consistent with previous studies \cite{luding11,luding08p,luding08crs,dijksman10,ries07, fenistein04,fenistein03}
 on cohesive and non-cohesive assemblies.
In Fig.\ \ref{fig:pos_coh}, the lines are the prediction by Unger et al.\ \cite{Unger04shear}:
\begin{equation}
z=H-R_c\left\{1-\frac{R_s}{R_c}\left[1-\left(\frac{H}{R_s}\right)^\beta\right]\right\}^{1/\beta}~,
\label{eq:Rc-z}
\end{equation}
where the exponent is given by $\beta=2.5$ for non-cohesive particles.
If the Bond number is less than one, the data collapse on a unique curve, very well
 predicted by Eq.\ \eqref{eq:Rc-z}, with fixed exponent $\beta$. On the other hand, 
above $Bo=1$, the exponent $\beta$ decreases with the global Bond number (values reported in Table \ref{tab:fitparameters}).
Note that Eq.\ \eqref{eq:Rc-z} slightly deviates from the results near the top surface if the cohesion is strong ($Bo=2.22$ and $2.85$).
In Fig.\ \ref{fig:wid_coh}, the lines are the prediction by Ries. et al.\ \cite{ries07} for non-cohesive system:
\begin{equation}
W(z)=W_{\rm top} \left[1-\left(1-\frac{z}{H}\right)^2 \right]^{\gamma}~,
\label{eq:W-z}
\end{equation}
where $W_{\rm top}$ is the width at the top surface and the exponent is given by $\gamma=0.5$ for non-cohesive particles.
If $Bo<1$, Eq.\ \eqref{eq:W-z} with $W_{\rm top}=0.012$ and $\gamma=0.5 \pm 0.1$ well agrees with our results.
However, for $Bo>1$, both the width $W_{\rm top}$ and exponent $\gamma$ increase with the global Bond number as in Table \ref{tab:fitparameters}.
In addition, Eq.\ \eqref{eq:W-z} deviates from the results near the top layer if the cohesion is strong ($Bo=2.22$ and $2.85$),
where $W$ seems to saturates above $z/H \simeq 0.6$.
Hence for $Bo>1$, we choose width at that height to be $W_{\rm top}$ and use $\gamma=0.66$ and $0.7$ for $Bo=2.22$ and $2.85$, respectively.

From the above results, we conclude that the cohesive forces between particles drastically affect the flow profiles.
Eqs.\ \eqref{eq:Rc-z} and \eqref{eq:W-z} very well predict the position and width of the shear bands for $Bo<1$.
 For large $Bo$ these equations deviate from observed behavior at large heights since the shear band interferes with the inner cylinder.
The shear band, which is the region with large velocity gradient, is caused by \emph{sliding motions} of particles.
However, strong cohesive forces keep particles in contacts
(in other words, the cohesive forces promote \emph{collective motions} of particles) and prevent them from sliding.
As a result, the velocity gradient is smoothened and the width of shear-band is broadened.
 This observation is consistent with previous studies on adhesive dense emulsions \cite{Ovarlez08}.
Interestingly, such an effect of cohesion is suppressed if the global Bond number is less than one,
where our numerical data agrees well with previous theoretical/numerical studies on non-cohesive particles \cite{Unger04shear,ries07}.
Hence, the global Bond number, $Bo$, captures the transition between essentially non-cohesive free-flowing granular assemblies
 $(Bo<1)$ to cohesive ones $(Bo>1)$.

\begin{table*}
\begin{tabular}{|l|l|l|l|l|l|l|l|}
\hline
    $Bo$ & $A_1$ & $A_2$ & $H$ & $\beta$ & $\frac{z}{H}$ range & $W_{\rm top}$ & $\gamma$ \\
\hline
    0 & 0.50$\pm$ 0.0005 & 0.500$\pm$ 0.0005 & 0.0365 & 2.52 & 0.1-1& 0.0117 & 0.507 \\
    0.17 & 0.50$\pm$ 0.0005 & 0.499$\pm$ 0.0005 & 0.0365 & 2.52 & 0.1-1 & 0.0118 & 0.523 \\
    0.33 & 0.49$\pm$ 0.0007 & 0.500$\pm$ 0.0007 & 0.0365 & 2.512 & 0.1-1 & 0.0118 & 0.555 \\
    0.81 & 0.49$\pm$ 0.0008 & 0.500$\pm$ 0.0008 & 0.0361 & 2.494 & 0.1-1 & 0.0119 & 0.583 \\
    1.05 & 0.49$\pm$ 0.001 & 0.501$\pm$ 0.001 & 0.0359 & 2.510 & 0.1-1 & 0.0120 & 0.582 \\
    1.50 & 0.49$\pm$ 0.002 & 0.501$\pm$ 0.002 & 0.0364 & 2.453 & 0.1-0.8 & 0.0126 & 0.613 \\
    2.22 & 0.49$\pm$ 0.003 & 0.501$\pm$ 0.003 & 0.0368 & 2.367 & 0.1-0.6 & 0.0138 & 0.667 \\
    2.85 & 0.49$\pm$ 0.005 & 0.502$\pm$ 0.005 & 0.0369 & 2.259 & 0.1-0.6 & 0.0160 & 0.713 \\
\hline
\end{tabular}
\caption{Table showing filling height of the system $H$, and fitting range ${z}/{H}$ for Eqs.\ \eqref{eq:Rc-z} and \eqref{eq:W-z},
together with the fit parameters $A_1$, $A_2$ in Eq.\ \eqref{eq:v_phi}, $\beta$ in Eq.\ \eqref{eq:Rc-z}, 
 $W_{top}$ and $\gamma$ in Eq.\ \eqref{eq:W-z} for different values of Bond number $Bo$.}
\label{tab:fitparameters}
\end{table*}
%
\subsection{Structure and distribution of forces in shear bands}
\label{sub:struc}
%
\begin{figure}
\includegraphics[scale=0.5]{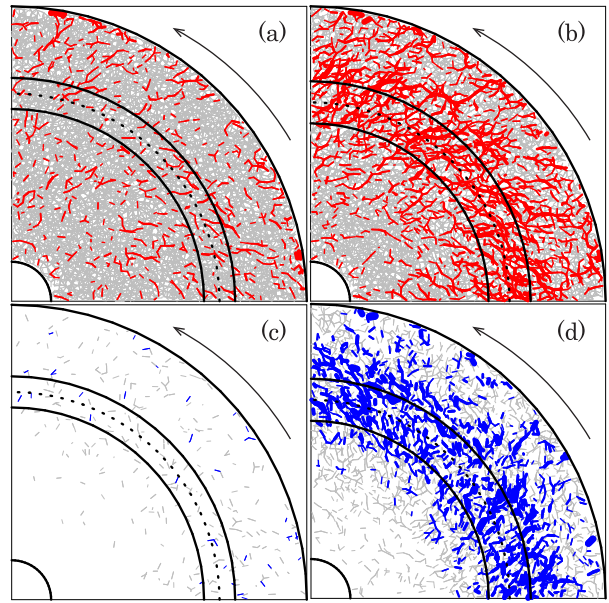}
\caption{(Color online)
Force chain networks of positive normal forces for $Bo=$ $0.33$ (a) and $2.85$ (b),
and negative normal forces for $Bo=$ $0.33$ (c) and $2.85$ (d) at height $0.02<z<0.025$\,m, respectively.
In (a) and (b) positive normal force smaller than $0.002$\,N is represented by grey, while larger than $0.002$\,N is represented by red color.
In (c) and (d) negative normal force smaller than $-0.0005$\,N is represented by grey, while larger than $-0.0005$\,N is represented by blue color.}
\label{fig:f_network}
\end{figure}
\begin{figure}
\mbox{
\subfigure{\includegraphics[scale=0.3,angle=-90]{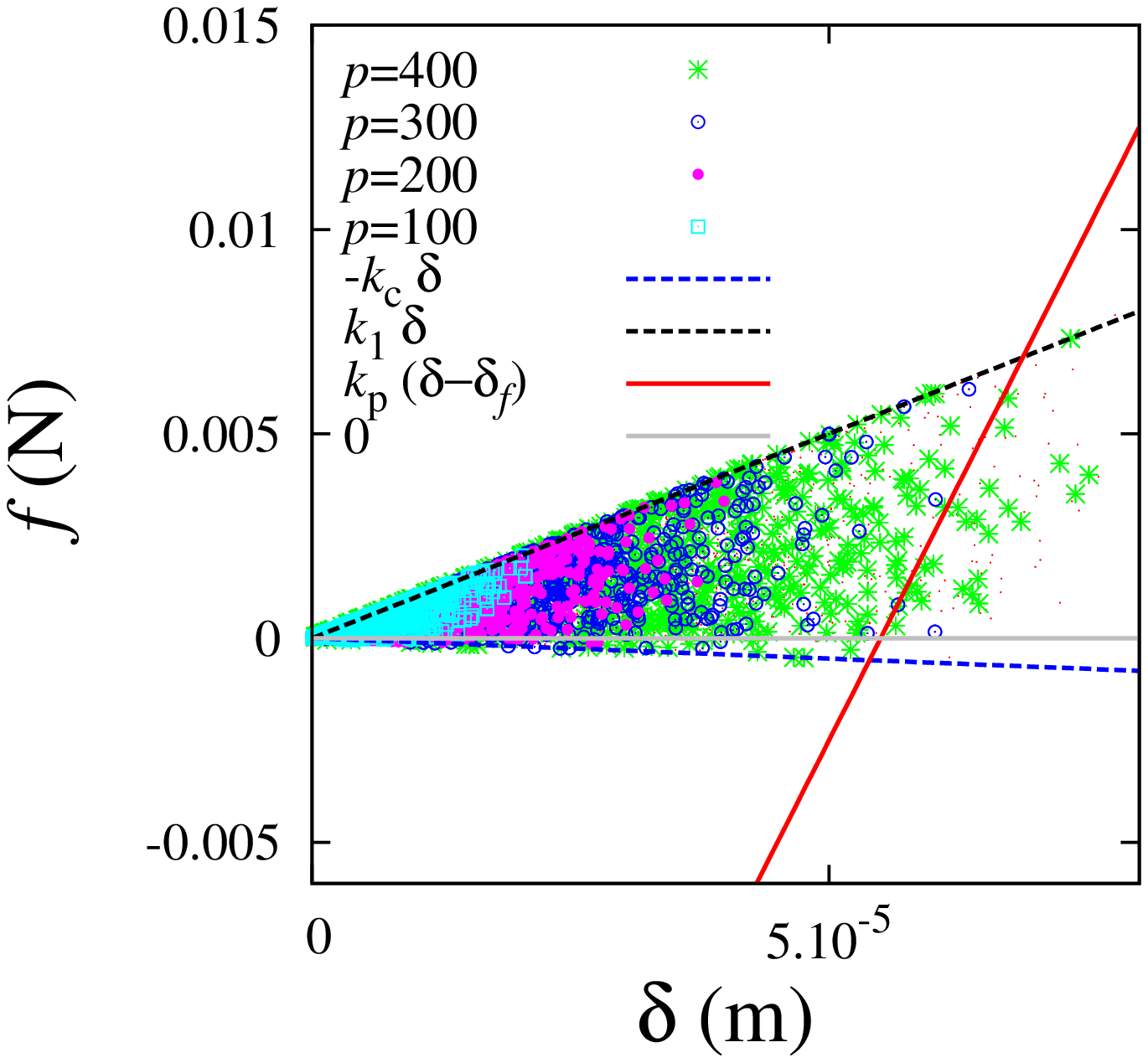}\label{fig:large_vel_stick}}
\subfigure{\includegraphics[scale=0.3,angle=-90]{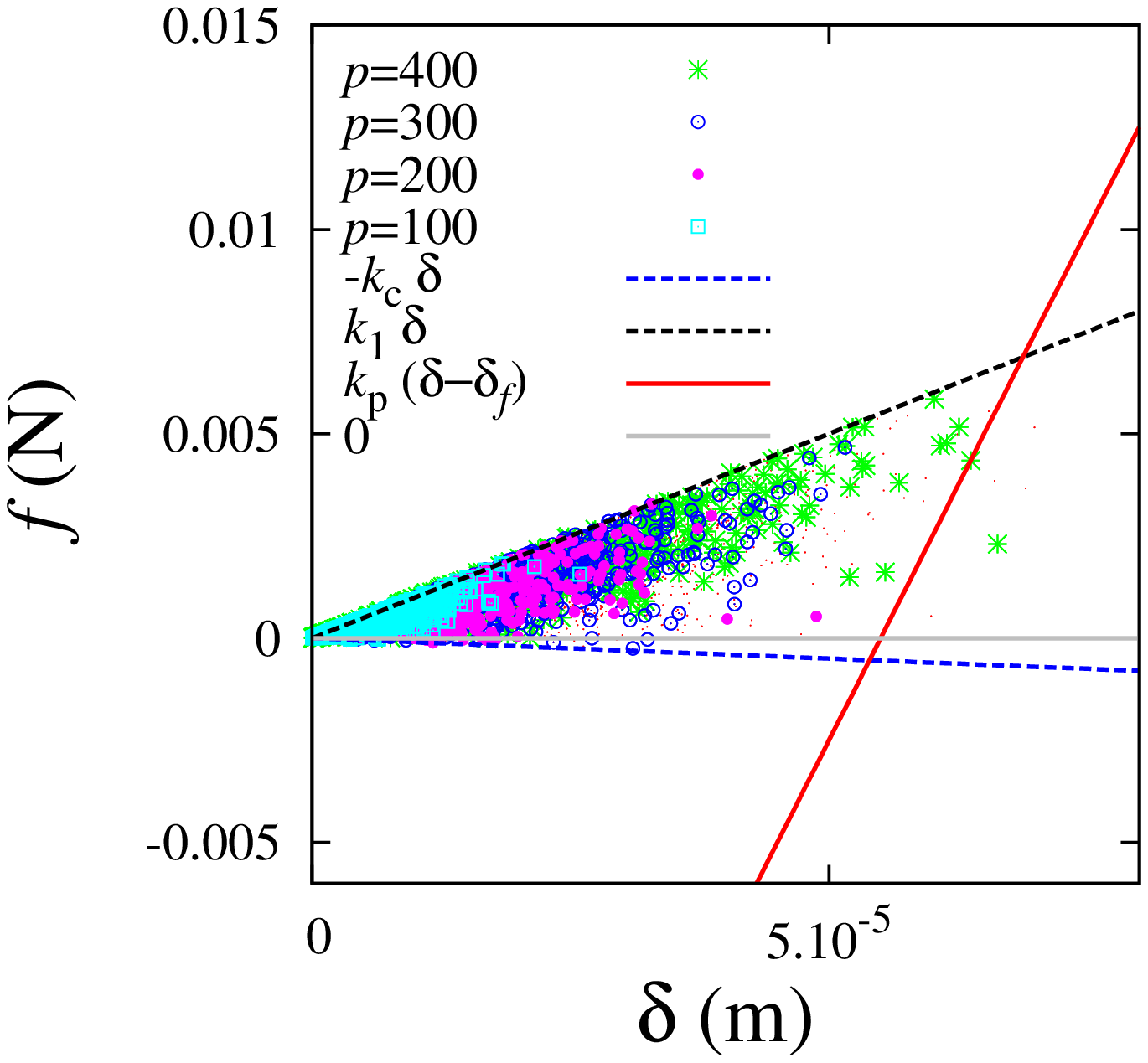}\label{fig:large_vel_force}}}
\mbox{
\subfigure{\includegraphics[scale=0.3,angle=-90]{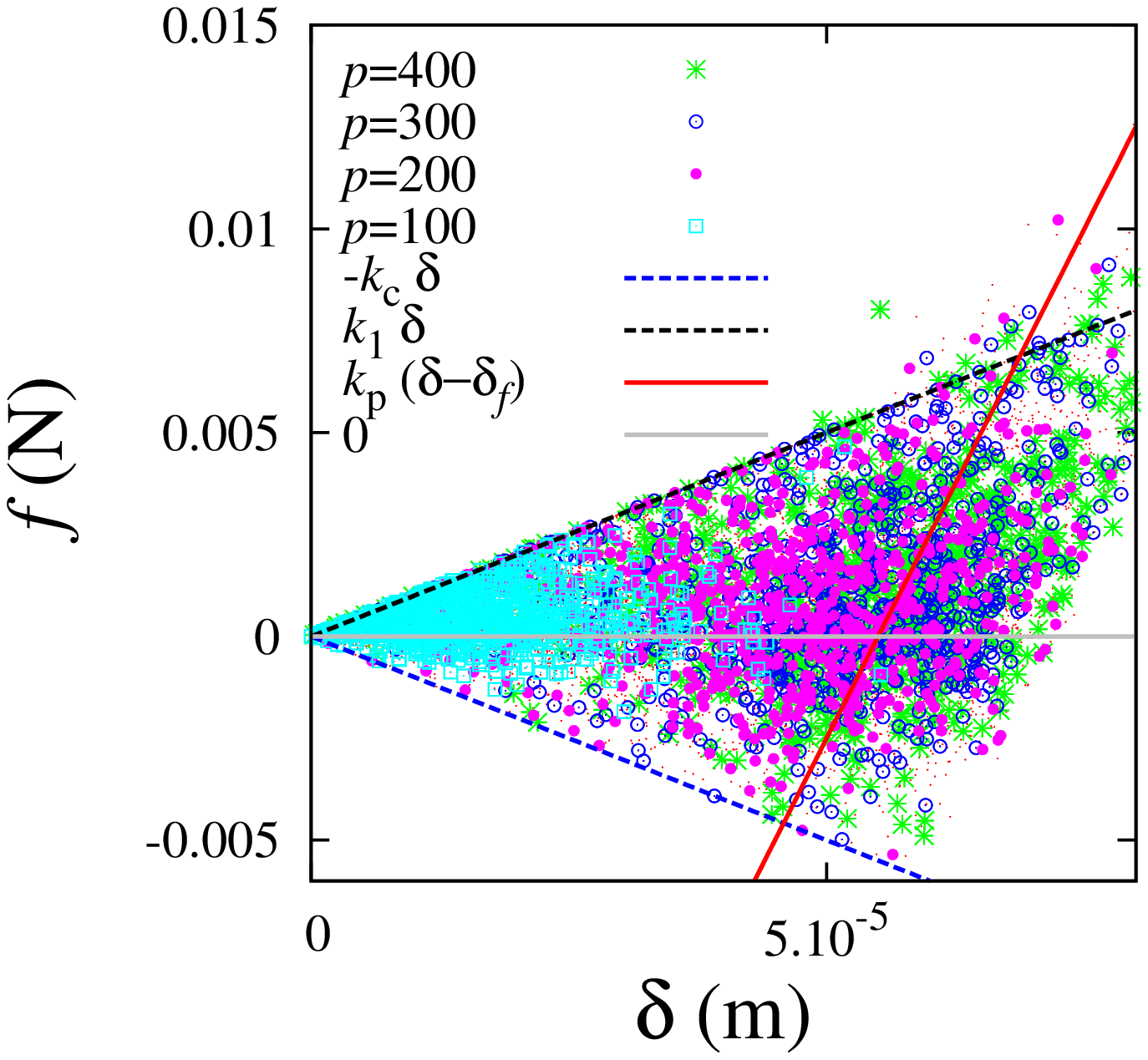}\label{fig:inside_coh}}
\subfigure{\includegraphics[scale=0.3,angle=-90]{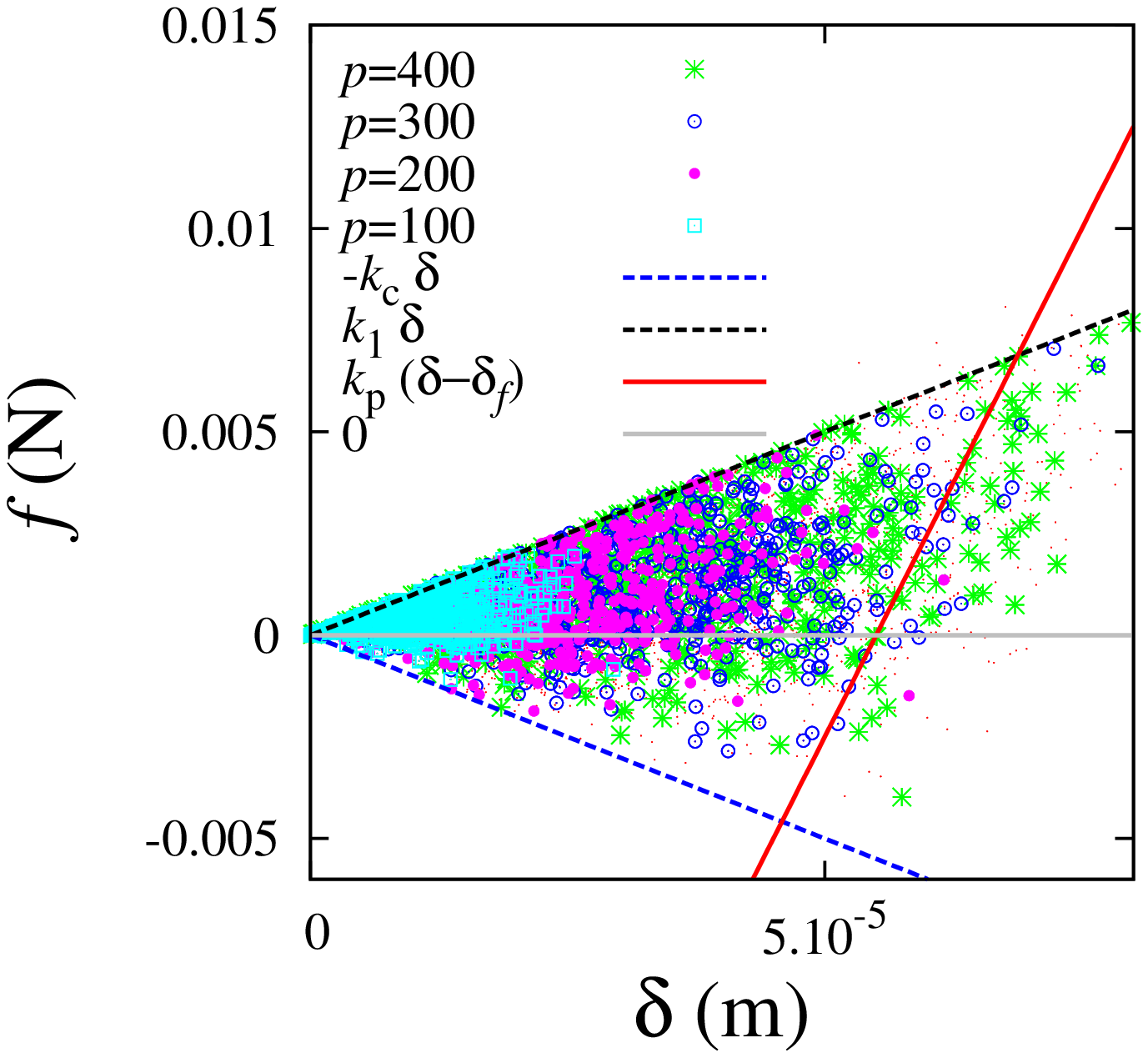}\label{fig:outside_coh}}}
\caption{(Color online)
Scatter plots of overlaps and forces between all contacts inside (left) and outside (right) of the shear bands for different $Bo=0.33$ and $2.85$.
The different symbols represent a zoom into the vertical ranges $z=8$\,mm $\pm 1$\,mm (green stars),
15\,mm $\pm 1$\,mm (blue circles), 22\,mm $\pm 1$\,mm (magenta dots), 29\,mm $\pm 1$\,mm (cyan squares),
with approximate pressure as given in the inset.
Note that the points do not collapse on the line $k_p(\delta-\delta_f)$ due to the finite width of the size distribution:
pairs of larger than average particles fall out of the indicated triangle.
Radial range $0.075$\,m\,$\le r \le 0.085$\,m (left) signifies data points inside the shear band,
while the radial range $0.055$\,m\,$\le r \le 0.065$\,m (right) signifies the data points outside the shear band.}
\label{fig:pcontacts}
\end{figure}
%
To understand the microscopic origin of the anomalous flow profiles of cohesive aggregates,
we study the force network and the statistics of the inter-particle normal forces. 
Recently Wang et al.\ \cite{ABYu13} reported the shape of probability distribution function (PDF) as an indicator for transition from quasistatic to inertial flows.
 In this section, we use a similar philosophy and study the change in the shape of PDFs as the cohesive strength is increased.

Figure \ref{fig:f_network} shows force chains of positive ((a) and (b)) and negative ((c) and (d)) normal forces
in the systems with low cohesion ((a) and (c)) and strong cohesion ((b) and (d)).
Grey color shows the weak forces, while red and blue colors show the strong positive and negative forces, respectively.
The strong or weak positive forces are forces larger or smaller than the mean positive force $f_{\rm pos}$. A similar
 approach is adopted to identify the strong/weak negative forces.
In this figure, we observe that both positive and negative forces are fully developed in the cohesive system ((b) and (d)),
with the intensity of the force inside the shear band being stronger than outside.
In addition, the strong (positive/negative) force chains are percolated through the shear band region.
As explained in Sec.\ \ref{sub:aniso}, we can also see that the positive and negative force chains
are aligned in their preferred directions,\ i.e.\ compressive and tensile directions, respectively.

Figure \ref{fig:pcontacts} displays scatter plots of the inter-particle forces against overlaps between the particles in contacts,
where each point corresponds to a contact and different colors represent different height,\ i.e.\ pressure level in the system.
The higher the pressure $p$, the higher is the average force (or overlap), as it must sustain the weight of the particles.
For almost all values of $Bo$, the density of points towards the unloading branch $k_p$ is higher inside the shear band compared to outside.
We also observe that with increasing $Bo$, most contacts (except for small pressure) drift towards
and collapse around the limit branch.
This implies that \emph{the cohesive forces are more pronounced in shear bands} rather than outside.

%
%
\subsubsection{Mean force and overlap in shear bands}
%

\begin{figure}
\includegraphics[scale=0.35,angle=-90]{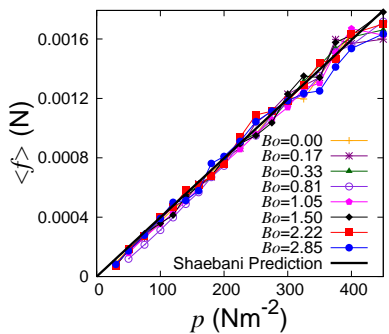}
\caption{(Color online)
The mean normal force $\langle f \rangle$ inside of the shear band plotted against pressure $p$,
where different symbols represent the global Bond number (as given in the inset)
and the solid line is given by Eq.\ (\ref{eq:f-P}).}
\label{fig:f_coh_comp}
\end{figure}

%
Figure \ref{fig:f_coh_comp} displays the mean normal forces, $\langle f \rangle$,
in the shear band plotted against pressure, for different values of the global Bond number,
where the solid line is the prediction by Shaebani et al.\ \cite{Shebani11} for non-cohesive granular systems:
\begin{equation}
\langle f \rangle = \frac{4\pi \langle a^2 \rangle }{\phi C g_2} \langle p \rangle
\label{eq:f-P}
\end{equation}
with the $2^{\rm nd}$ moment of the size distribution $\langle a^2\rangle$, coordination number $C$, volume fraction $\phi$, and mean pressure $\langle p \rangle$.
Notably, \emph{the mean normal force is almost independent of cohesion} and linearly increases with pressure
as in the cases of static non-cohesive \cite{mueth98,silbert2002statistics} and cohesive systems \cite{ABYucohfor08}.
We also observe that for low pressure, Eq.\ \eqref{eq:f-P} slightly over predicts the value of the mean force, while for higher pressure the prediction 
well captures the data.
\begin{figure}
\subfigure[]{\includegraphics[scale=0.35,angle=-90]{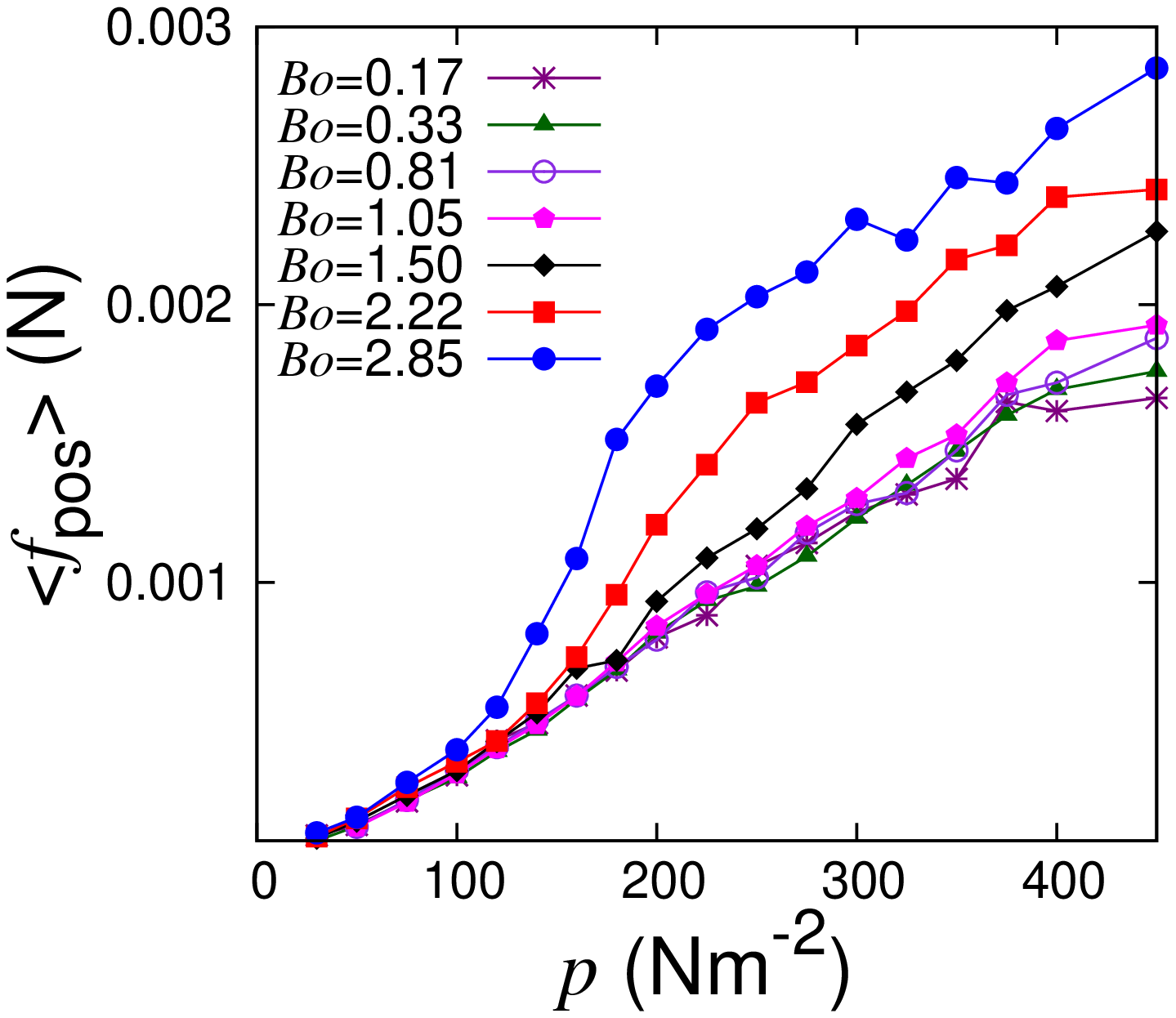}}\\
\subfigure[]{\includegraphics[scale=0.35,angle=-90]{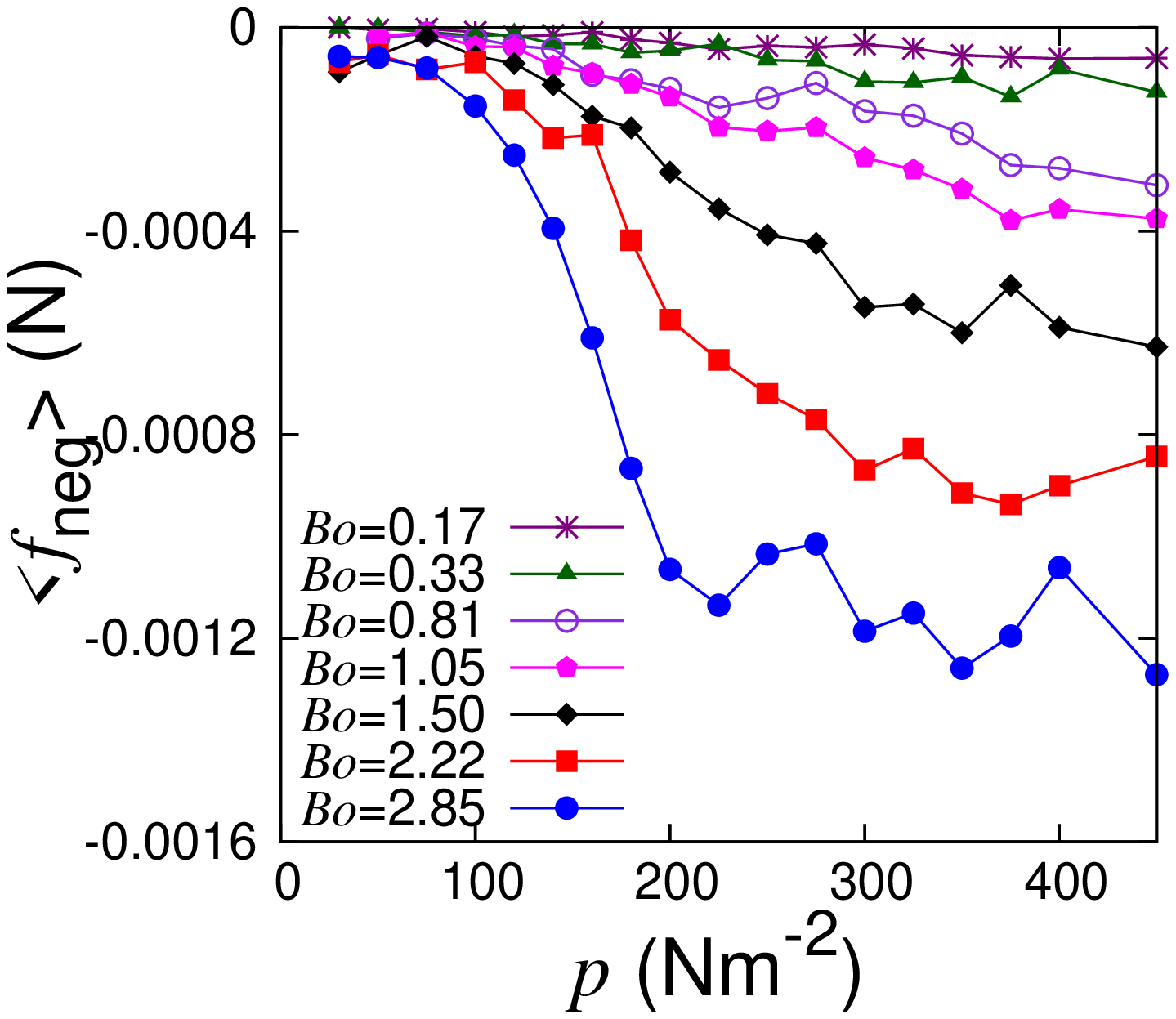}}
\caption{(Color online)
The mean (a) positive force $\langle f_{\rm pos} \rangle$ and (b) negative force $ \langle f_{\rm neg} \rangle$ inside the shear band plotted against pressure $p$,
where different symbols represent the global Bond number  (as given in the inset).}
\label{fig:fpm}
\end{figure}
%
While the mean value is insensitive to cohesion, 
the mean positive and negative normal forces, $\langle f_{\rm pos} \rangle$ and $ \langle f_{\rm neg} \rangle$, strongly depend on cohesion,
 we plot them in Fig.\ \ref{fig:fpm} against pressure for different values of $Bo$,
where \emph{the intensities increase with cohesion} in agreement with Fig.\ \ref{fig:f_network}.
Note that the mean positive (negative) force is linear with pressure and independent of cohesion below $Bo=1$,
while its dependence on pressure becomes nonlinear above $Bo=1$.
Though the origin of this nonlinearity is not clear, it is readily understood that cohesion 
enhances the collective motion of the particles,\ i.e.\ the particles rearrange less and the system is in a mechanically constrained state.
Such constrain leads to increase in the magnitude of negative forces.
 As a consequence, positive forces also increase, in order to balance the negative ones.
It is noteworthy that in Fig.\ \ref{fig:pcontacts}, the increase of $Bo$ increases the density of points
in both positive and negative extremes, inside the shear band, in accordance with the previous
 considerations.

\begin{figure}
\subfigure[]{\includegraphics[scale=0.35,angle=-90]{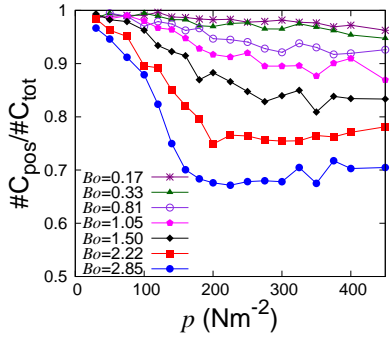}}\\
\subfigure[]{\includegraphics[scale=0.35,angle=-90]{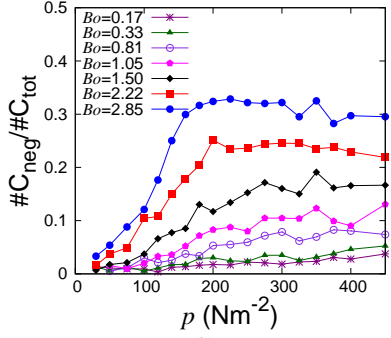}}
\caption{(Color online)
The fractions of (a) positive and (b) negative contacts inside the shear band plotted against pressure $p$,
where different symbols represent the global Bond number  (as given in the inset).}
\label{fig:cpm}
\end{figure}
%

Similar to what observed for the mean force, cohesion seems not to affect the average number of contacts,
 as reported in Ref.\ \cite{asingh13}, where we observed that cohesion had practically no effect on the contact number density
 (volumetric fabric). Fig. \ref{fig:cpm} shows the fractions of repulsive and attractive contacts against pressure for different Bond numbers,
 together with the overall coordination number. 
An increase of cohesion generates more attractive contacts while the number of repulsive contacts decrease.
 Interestingly, the overall mean force remains independent of cohesion and contacts
 simply redistribute between the repulsive and attractive directions.

\begin{figure}
\includegraphics[scale=0.35,angle=-90]{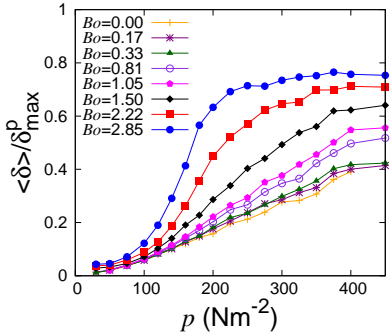}
\caption{ (Color online)
Normalized mean overlap $\frac{<\delta>}{\delta^p_{\rm max}}$ inside the shear band plotted against pressure $p$,
where different symbols represent the global Bond number (as given in the inset).}
\label{fig:overlap_coh_comp}
\end{figure}

In contrast to the mean force, the mean overlap between particles
 in contact depends on cohesion non-linearly, as shown in Fig.\ \ref{fig:overlap_coh_comp}. In our model of cohesive particles \cite{Luding08gm},
 overlaps are always positive for both positive and negative forces.
It is worth mentioning that for low $Bo$, $\langle \delta(t) \rangle$ saturates quickly,
while for $Bo=1.5,2.22,2.85$ it takes longer to reach the steady state due to the average plastic increase of the overlap \cite{luding11}.

%

%
\subsubsection{PDFs of forces and structures of strong force chains in shear bands}
%
\begin{figure}
\subfigure[]{\includegraphics[scale=0.35,angle=-90]{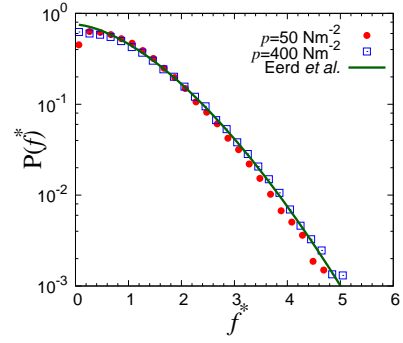}\label{fig:kc0-dist1}}\\
\subfigure[]{\includegraphics[scale=0.35,angle=-90]{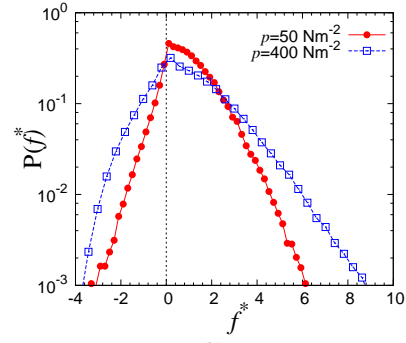}\label{fig:kc100-dist}}
\caption{(Color online)
Probability distribution of the normalized force $f^*$ for (a) cohesion-less $Bo=0$
and (b) highly cohesive $Bo=2.85$ systems at different pressures $p$ in the system.
Different symbols represent value of local pressure (as given in the inset).}
\label{fig:kc-dist}
\end{figure}
\begin{figure}
\subfigure[]{\includegraphics[scale=0.35,angle=-90]{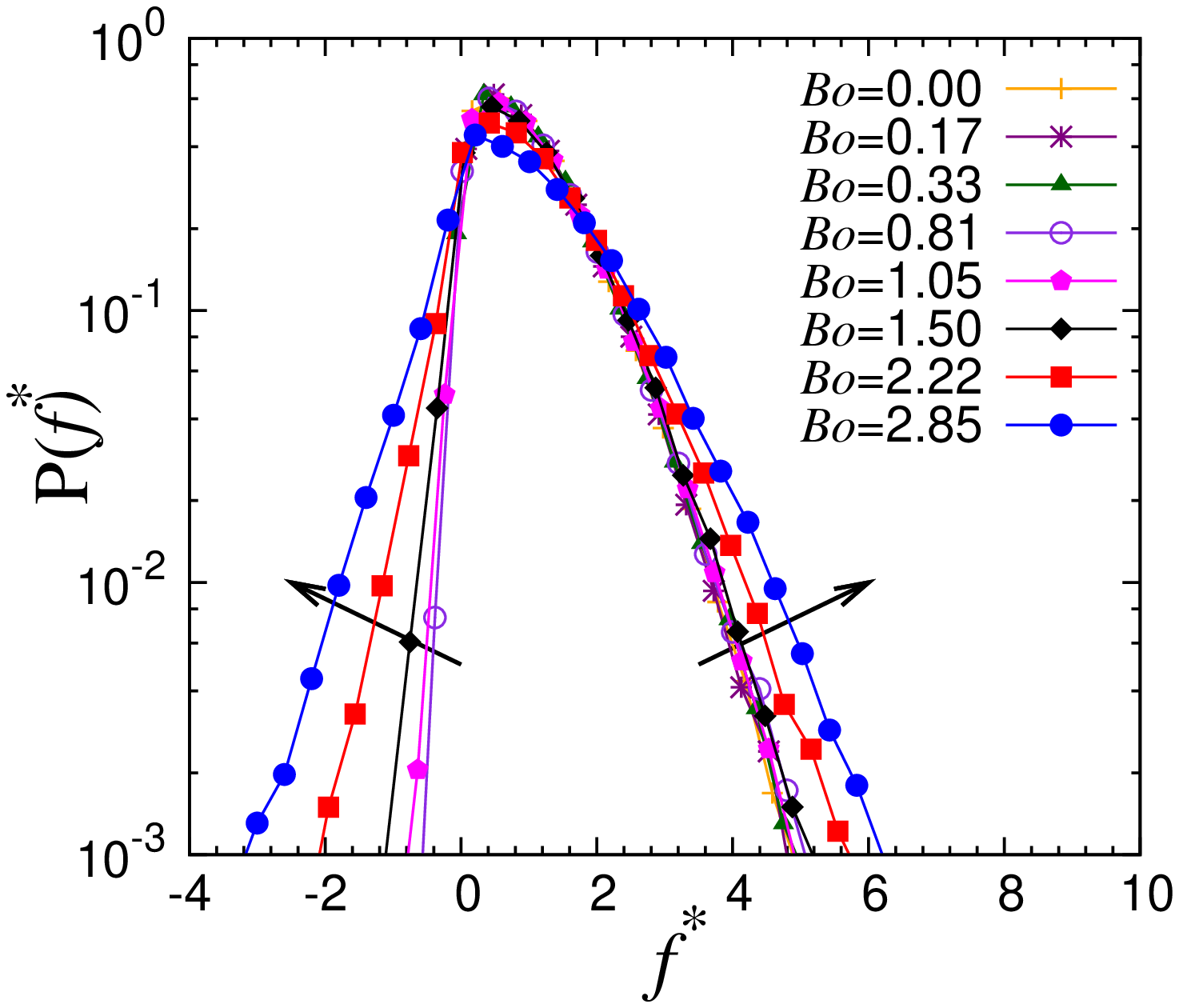}\label{fig:pdf-force-lowp}}\\
\subfigure[]{\includegraphics[scale=0.35,angle=-90]{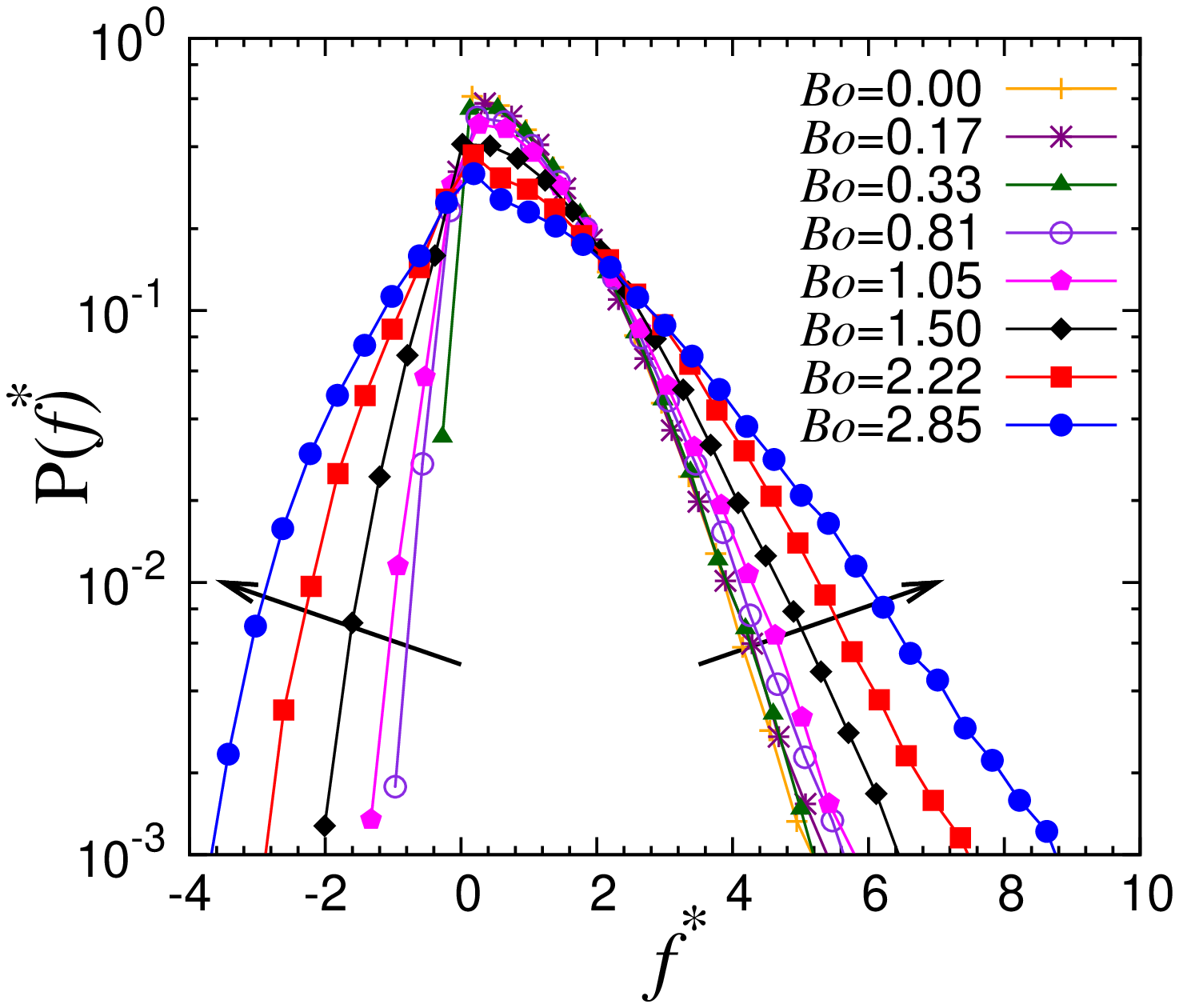}\label{fig:pdf-force-highp}}
\caption{(Color online)
Probability distribution of normalized force $f^*$ for (a) low pressure $p=50$\ $\mathrm{Nm^{-2}}$ (close to top)
and (b) high pressure $p=400$\ $\mathrm{Nm^{-2}}$ (close to bottom) in the system for data inside the shear band.
Different symbols represent the global Bond number $Bo$ (as given in the inset).}
\label{fig:pdf-force_coh_comp}
\end{figure}
%
The probability distribution function (PDF) of forces are also strongly affected by cohesion.
Figure \ref{fig:kc-dist} shows the PDFs of normal forces in shear bands for different pressure and cohesion,
where the forces are scaled by the mean normal force,\ i.e.\ $f^\ast\equiv f/\langle f \rangle$.
As can be seen, the PDF of cohesion-less particles ($Bo=0$) is almost independent of pressure (Fig.\ \ref{fig:kc0-dist1}),
while it \emph{depends on pressure} if the cohesive forces are very strong (Fig.\ \ref{fig:kc100-dist}).
Figure \ref{fig:pdf-force_coh_comp} displays the variations of the PDFs for different intensities of cohesion,
where we find that the PDF becomes broad with increasing cohesion and $Bo>1$.
Therefore, \emph{ strong cohesion, which leads the system to a \textquotedblleft mechanically frustrated state\textquotedblright \ induces larger
 fluctuations of positive/negative forces}.
We note that Yang et al.\ \cite{ABYucohfor08} also found similar trends in static three-dimensional packing for small sized particles,
where the PDF becomes broader, as particle size decreases, i.e. cohesion increases.
Broadening of the PDFs was also observed by Luding et al.\ \cite{Luding2005} during cooling down of a sintered system.

\begin{figure}
\subfigure[]{\includegraphics[scale=0.35,angle=-90]{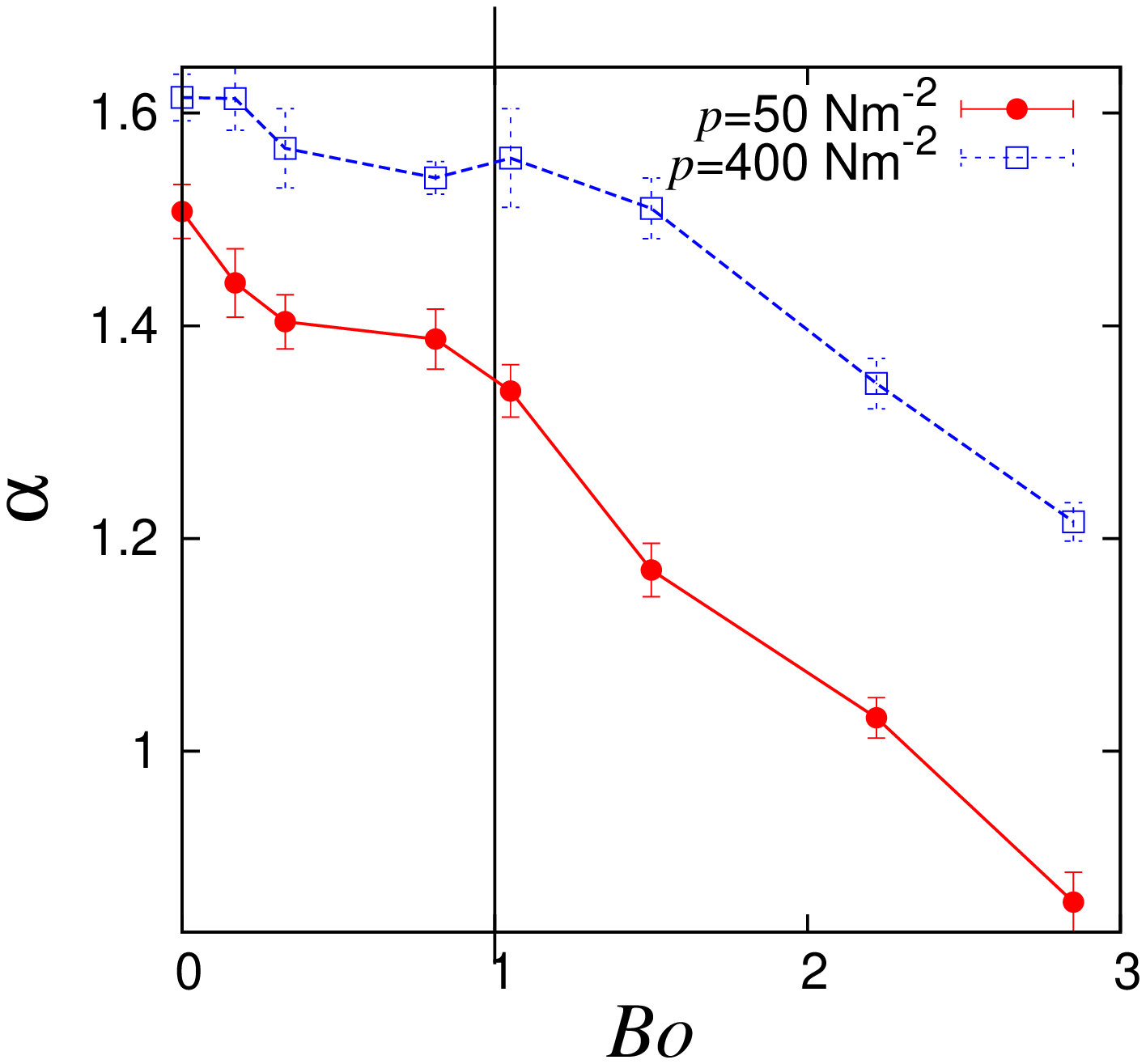}\label{fig:alpha_p}}
\subfigure[]{\includegraphics[scale=0.35,angle=-90]{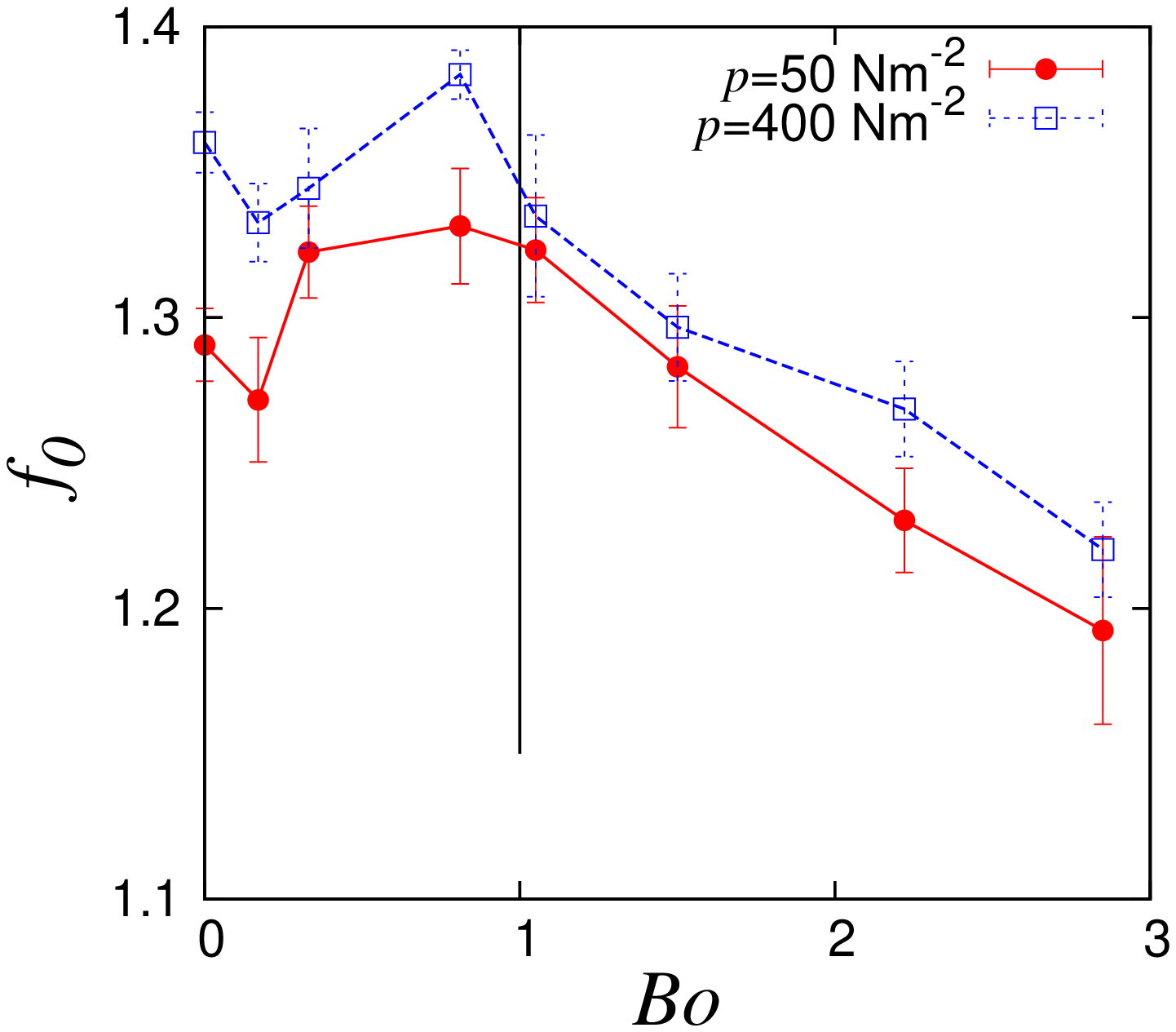}\label{fig:fc_p}}
\caption{ (Color online)
Fit parameters (a) $\alpha$ and (b) $f_0$ plotted against Bond number $Bo$.
Different symbols represent value of local pressure (as given in the inset).}
\label{fig:log-pdf-force_coh_comp}
\end{figure}
%
The cohesive forces modify not only the shapes of the PDFs, but also their asymptotic behavior,\ i.e.\ the structure of strong force chains.
The tails can be fitted by a stretched exponential function \cite{Hecke_tail05}
\begin{equation}
P(f^*) \sim e^{-(f^*/f_0)^\alpha}
\label{eq:dist-f}
\end{equation}
with a characteristic force $f_0$ and a fitting exponent $\alpha$.
Figure \ref{fig:log-pdf-force_coh_comp} displays the characteristic force and the exponent against the global Bond number $Bo$.
If $Bo<1$, we obtain $f_0=1.4\pm0.1$ and $\alpha=1.6\pm0.1$, which is very close to that predicted by Eerd et al.\ \cite{Hecke_tail05}
for three-dimensional non-cohesive ensemble generated by MD simulations.
However, for $Bo>1$, both characteristic force and fitting exponent decrease with increasing cohesion.
The decreasing fitting exponent hints at stronger fluctuations in the force distribution.
 A Gaussian tail of the probability distribution would indicate a more homogeneous random spatial distribution of forces.
 The deviation towards an exponential distribution can be linked to an increase in heterogeneity in the spatial force distribution
 as mentioned in previous studies \cite{radjai1996force,makse2000packing,zhang2005jamming}.
Therefore, we conclude that 
\emph{the tail of the PDF becomes a wider exponential with increasing cohesion,
which implies a more heterogeneous spatial distribution, especially of the strong forces}.

Finally we observe that the fitting exponent decreases with increasing pressure,
which implies that at high pressure where cohesion is more active due to the contact model the spatial distribution is more heterogeneous compared to low pressure.
%
%
\subsection{Anisotropy of force chain networks in shear bands}
\label{sub:aniso}
%
\begin{figure}
\includegraphics[scale=0.4]{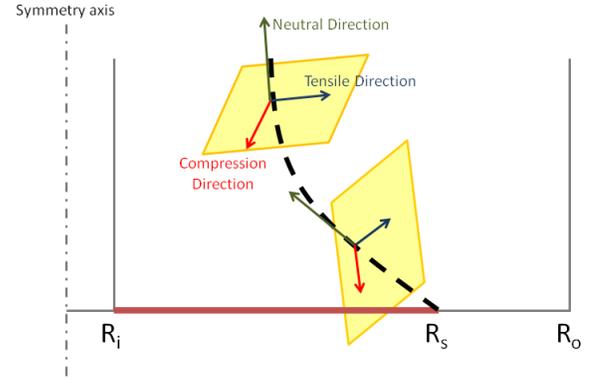}
\caption{(Color online)
A sketch showing the shear band, shear plane, and three eigen-directions of the strain rate tensor. Grey lines show inner and outer cylinders,
 while solid brown line shows the split, dashed black line shows the shear band which initiates at the split at bottom and moves towards inner cylinder as it moves
 towards the top. Green arrow represents the eigen-direction for neutral eigenvalue of the strain rate tensor, which is tangential to the shear band, perpendicular
 to this vector is the shear plane (yellow shaded region), which contains the eigen-directions for compression (red arrow) and tensile (blue arrow) eigenvalues. $R_i$,
 $R_s$ and $R_o$ show the inner, split, and outer radii respectively.}
\label{fig:sketch}
\end{figure}
%
In the case of simple shear as developed in the split-bottom shear cell, there are two non-zero eigenvalues of the strain rate tensor,
which are equal in magnitude but opposite in sign, while the third eigenvalue is zero.
The plane containing the eigen-vectors associated to non-zero eigenvalues is called the \textquotedblleft shear plane\textquotedblright,
and the eigen-vector with zero eigenvalue is perpendicular to this plane (tangent to the shear band).
In the following we will refer to the eigen-directions associated to positive, negative, and zero eigenvalues as {\em compressive}, {\em tensile}, and {\em neutral} directions, respectively.

Note that the shear band here is not vertical, instead its orientation changes with depth as shown by the schematic in Fig.\ \ref{fig:sketch}.
In this figure, the eigen-direction of the neutral (zero) eigenvalue (green arrow) moves with the shear band.
This turning of the neutral eigen-direction makes the shear plane tilt as well (which is shown by the yellow shaded regions).
To extract the contacts aligned along these directions at a given pressure in the system, we first calculate the local tensor at a given strain-rate and extract
 the three eigen-directions $\mathbf{n}^{\rm i}_\mathrm{\gamma}$ (with $i$ being compressive, neutral and tensile).
Next, we look for contacts with unit contact vector $\mathbf{n}_\mathrm c$, which satisfy the condition $ {|\mathbf{n}_\mathrm c\cdot {\mathbf{n}^{\rm i}_{\gamma}} |}$ $ \ge 0.9$ .
The contacts which satisfy the condition for compressive eigen-direction are termed compressive; tensile and neutral contacts are defined in a similar fashion.
The forces carried by compressive, tensile, and neutral contacts are denoted by $f_{\rm com}$, $f_{\rm ten}$, and $f_{\rm neu}$ respectively.

Since compressive and tensile directions are associated with loading and unloading of contacts, respectively,
it is intuitive that in the absence of any external force other than shear, the mean force would be positive in compressive direction,
negative in tensile direction, and almost zero in neutral direction.

In our system, both an external load -- gravity-- coexist with (external) shear.
 The neutral direction gets a contribution from the additional load only,
while the two principal (compressive and tensile) directions get contributions from both shear and gravity.
Because the cohesive force is activated by unloading, we expect that it affects the forces along the tensile direction.

\begin{figure}
\includegraphics[scale=0.4,angle=-90]{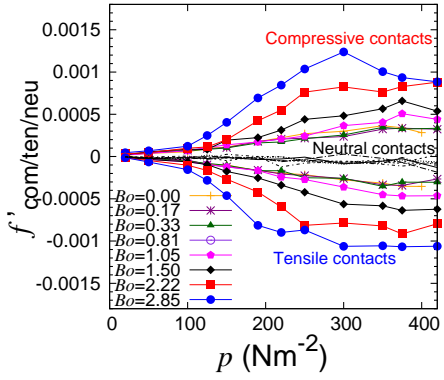}
\caption{(Color online)
Mean forces in different eigen-directions of the strain rate tensor, relative to
the overall mean force plotted against the local pressure $p$ in the system.
Different symbols represent the global Bond number $Bo$ (as given in the inset).}
\label{fig:f_coh_comp-dir}
\end{figure}
%

Figure \ref{fig:f_coh_comp-dir} shows the mean compressive/tensile/neutral forces relative to overall local mean force,
$ f^{'}_{\rm{com/ten/neu}}\equiv\langle f_{\rm{com/ten/neu}}\rangle - \langle f\rangle$, plotted against pressure for different values of $Bo$.
 We find that $ f^{'}_{\rm com}(>0)$ and $ f^{'}_{\rm ten}(<0)$ are symmetric about zero, and $ f^{'}_{\rm neu}\simeq 0$.
The mean force along the neutral direction is independent of $Bo$, as the cohesion does not affect $f_{\rm neu}$ due to the absence of shear in this direction.
However, $ f^{'}_{\rm ten}$ decreases with pressure and cohesion.
 At the same time $ f^{'}_{\rm com}$ increases in order to keep the mean overall force independent of cohesion.
 We point out here that the difference between positive and negative values, i.e., 
\emph{the anisotropy of forces becomes more pronounced with increasing pressure and cohesion}. This is consistent
 with the visual observation of force chains of negative and positive forces for different intensity of cohesion,
 as shown in Fig.\ \ref{fig:f_network}.

\begin{figure}
\subfigure[]{\includegraphics[scale=0.35,angle=-90]{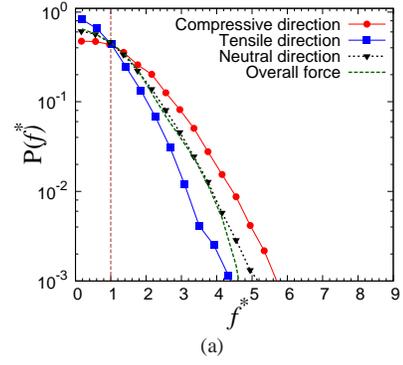}\label{fig:pdf-force-kc0-all}}\\
\subfigure[]{\includegraphics[scale=0.35,angle=-90]{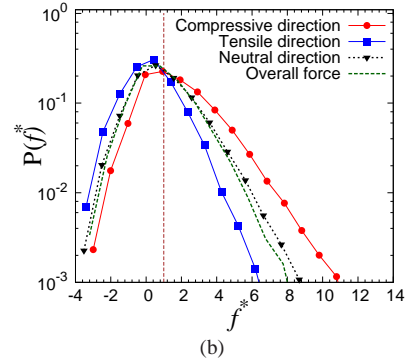}\label{fig:pdf-force-kc100-all}}
\caption{(Color online)
Probability distributions of normalized forces $f^\ast=f/\langle f \rangle$ 
in compressive, tensile, and neutral directions inside the shear bands for high pressure in (a) non-cohesive $Bo=0$ 
and (b) highly cohesive $Bo=2.85$ systems. The dashed line curves show the PDFs of the overall normalized forces,
while the average force is indicated by the vertical lines.}
\label{fig:pdf-force-kc-all-dir}
\end{figure}
%
Next, we study the PDFs of forces in the compressive, tensile, and neutral directions.
Figure \ref{fig:pdf-force-kc-all-dir} displays the PDFs along each direction for non-cohesive $Bo=0$ and highly cohesive $Bo=2.85$ systems,
where the forces along different directions are normalized by the overall mean force.
In a non-cohesive system (Fig.\ \ref{fig:pdf-force-kc0-all}),
we observe that for weak forces, i.e., $f^*<1$, the PDF along the tensile direction is higher compared to that for the compressive direction.
This is intuitive as for the weak forces, the majority of contacts will be aligned along the tensile direction.
However, for $f^*>1$ the PDF along the compressive direction becomes higher compared to that along the tensile direction,
as majority of contacts along the compressive direction should carry strong forces \cite{Brian05force}.
For a highly cohesive system (Fig.\ \ref{fig:pdf-force-kc100-all}), a similar behavior is observed for strong positive forces $f^*>1$.
While for weak positive and whole range negative forces, the PDF along the tensile direction is higher in comparison 
to the compressive direction.
The PDFs of forces in the neutral direction lie in between those in compressive and tensile directions,
suggesting a close to average distribution of forces.
It is interesting to note that, both positive and negative forces are present in all directions.
 However, the positive and negative forces dominate in the compressive and tensile directions, respectively.
\begin{figure}
\subfigure[]{\includegraphics[scale=0.35,angle=-90]{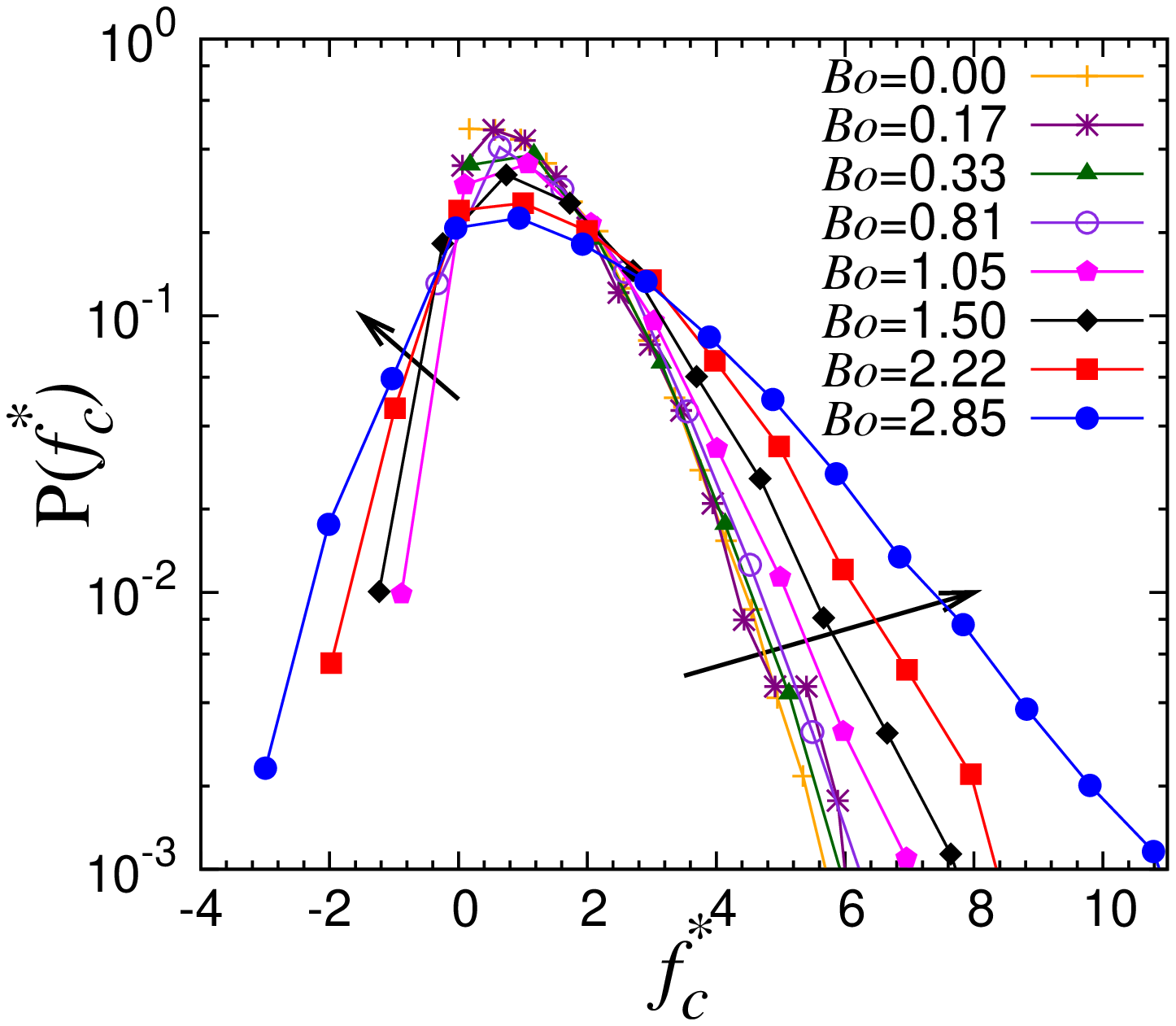}\label{fig:pdf-force-kcall-comp1}}\\
\subfigure[]{\includegraphics[scale=0.35,angle=-90]{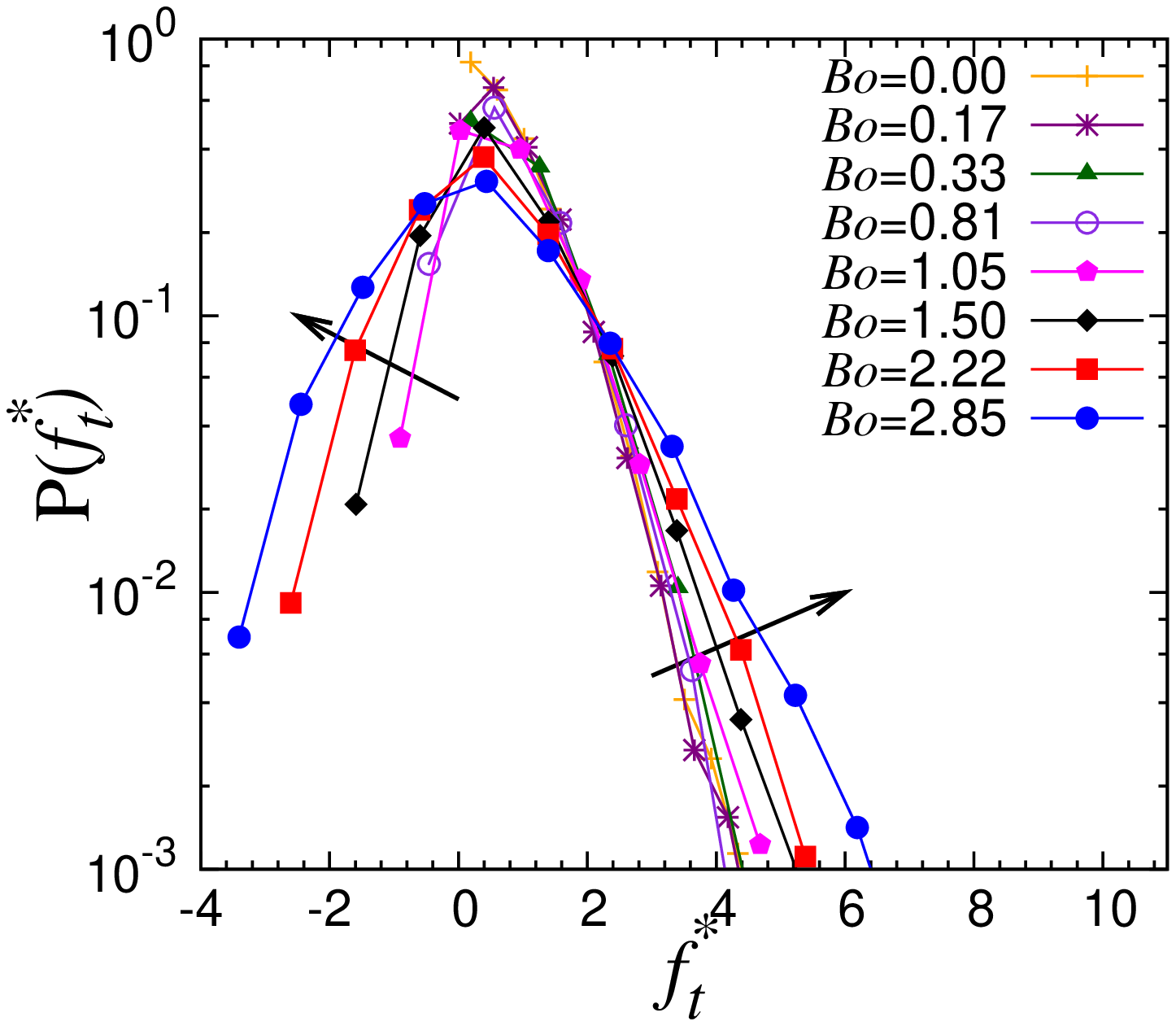}\label{fig:pdf-force-kcall-ten1}}
\caption{(Color online)
Probability distributions of normalized forces in (a) compressive ($f^\ast _c=f_c/\langle f \rangle$) 
and (b) tensile ($f^\ast _t=f_t/\langle f \rangle$) directions inside the shear bands for high pressure.
 Different symbols represent different values of the global Bond number $Bo$ as given in the inset.
}
\label{fig:pdf-force-kcall-dir}
\end{figure}
%

Figure \ref{fig:pdf-force-kcall-dir} shows the variations of the PDFs along compressive and tensile directions for different values of $Bo$.
If $Bo<1$, the PDFs collapse on top of each other.
However, the PDFs get wider with increasing cohesion above $Bo=1$.
Such widening is more prominent for positive and negative forces in the compressive and tensile directions, respectively.
Again, we confirm that strong cohesion leads to an increases of positive and negative forces in the compressive and tensile directions, respectively.
Therefore, as \emph{the force distributions in the principal directions gets more heterogeneous with increasing cohesion for $Bo>1$},
 the heterogeneity of the overall force structure increases.

Results in this section suggest that for low $Bo$, external load and shear dominate and govern the distribution 
of forces along compressive and tensile directions.
The forces can adapt to external shear, the particles rearrange and can avoid very large forces.
In contrast, for high $Bo$, cohesion dominates over external forcing: the contact forces still respond to compression
and tension, but their rearrangements are hampered by cohesion.
Due to the sticky nature of cohesive forces, rearrangements of the contact network become more difficult, so that
very large contact forces as well as strong sticking forces occur together, leading to a  more heterogeneous contact network.
%
%
\section{Discussion and conclusion}\label{sec:dis}
%
%
In this paper, we have studied the effect of cohesion on shear banding in dry cohesive powders.
The global {\em Bond number}, $Bo$, can be used to quantify how strong cohesive forces 
are relative to compressive forces, where $Bo \simeq 1$ very well predicts the transition from 
a free-flowing, non-cohesive system to a cohesive system.
Interestingly, many other features of the system also show a transition at $Bo\approx 1$. Using local
 $Bo$ has no big advantage in this system, but is recommended in general.

\paragraph{Shear band}
Width and center position of the shear band are fairly un-affected by cohesion for $Bo<1$;
only for $Bo \ge 1$ cohesion affects the flow behavior. 
The width of the shear band increases with $Bo$ increasing above unity; the velocity
gradient, as a consequence, decreases, since cohesive forces tend to keep the particles in contact 
to remain connected for longer. Cohesive forces assist the \textquotedblleft collective motion\textquotedblright\ 
of particles; implying that attractive forces work against the localization of shear.

\paragraph{Forces and their direction dependence}
The mean force $\langle f \rangle(p)$ (with $p\propto H-z $) is found to be independent of cohesion, 
just like the number of contacts.
With increasing $Bo$, stronger attractive negative forces are possible
 at contacts (which is intuitive). However, these negative forces must be balanced 
by some stronger positive forces to maintain the same overall mean force.

Due to the planar shear that establishes in steady state, compressive/tensile contact forces 
are induced in compressive/tensile eigen--directions of the local strain rate tensor, respectively, 
while along the third, neutral direction neither compression nor tension takes place. 
The mean force along the neutral direction remains unaffected by cohesion,
which implies that cohesive forces in the system are activated by shear; more
specific, cohesive forces are activated by the tension in the respective (eigen) direction. 
In other words, only about one third of all contacts features considerable strain-induced cohesion.

The mean force carried by contacts along compressive and tensile directions is {\em symmetric} about 
the mean overall force. For $Bo\le 1$, this anisotropy of the force network is independent of cohesion, 
while for $Bo> 1$ the anisotropy in the force network increases with cohesion.
Macroscopically, this anisotropy in force is directly related to the shear stress; 
the trend in force anisotropy is very similar to the trends found in the shear stress 
in previous work \cite{luding11}.

\paragraph{Force probability distribution}
Since granular systems are known to be heterogeneous in nature, we also analyzed the 
effect of cohesion on the force probability distributions.  For non-cohesive and
weakly cohesive systems, 
no prominent effect of pressure on force distributions could be seen. 
For strong cohesion $Bo>1$, pressure affects the distribution of forces, 
by making the tails wider, and more symmetric, as compared to the cases with $Bo< 1$.
Splitting up the force distributions along the compressive and tensile directions reveals 
that, 
for higher $Bo$, cohesion broadens the force distributions along the tensile direction, 
which in turn affects the distribution along the compressive direction, which also 
becomes wider. This suggests, an increase in heterogeneity in forces for $Bo>1$ 
along all directions, compressive, tensile and neutral as well. 
For low $Bo$, the kinematics of shear helps the particles to rearrange and 
avoid very strong forces. In contrast, for high $Bo$, cohesion induces stickiness 
at the contacts so that rearrangements are suppressed, increasing the heterogeneity 
of the system, as evidenced by the wider tails of the probability distributions.\\

In conclusion, both the flow profiles (shear banding) and the force structure are unaffected 
by cohesion for $Bo< 1$.  In contrast, for $Bo\ge1$, cohesion strongly affects the flow behavior, 
the anisotropy, and the internal force structure. Attractive forces thus reduce shear
localization for $Bo>1$ and promote heterogeneity of the force-network. 
These two observations are consistent with previous studies with attractive forces, 
concerning the rheology \cite{Chaudhuri2012} and force structures for static 
packings \cite{ABYucohfor08}. 

As speculation, for a wider view, our results can be interpreted as follows:
In the language of statistical mechanics, the global $Bo$ corresponds to a \textquotedblleft 
control parameter\textquotedblright\ and $Bo=1$ to a \textquotedblleft critical point\textquotedblright.
The changes in the characteristic force and the fitting exponents show a weak 
pressure dependence, which might be better captured using a pressure dependent, local Bond number. 
In our case, the macroscopic properties (position and width of shear-bands), the anisotropy and the micro-structural 
signatures (the tails of the PDFs) gradually increase for $Bo \ge 1$.
This continuous increase implies a \textquotedblleft second-order transition\textquotedblright,
however, confirming this would need a further detailed study. 
Also, experiments performed with controlled, pressure-dependent cohesive strength would 
be exciting to confirm and validate our results.
Finally it would be interesting to reproduce our findings with different contact models, 
e.g. capillary bridges or simpler cohesive contact models with no pressure dependence.

\begin{acknowledgments}\label{sec:acknow}

We thank M. Wojtkowski, N. Kumar, O. I. Imole, T. Weinhart, and Ruud van Ommen for 
 stimulating discussions. Financial support (project number: 07CJR06) through the 
``Jamming and Rheology'' program of the Stichting voor 
Fundamenteel Onderzoek der Materie (FOM), 
which is financially supported by the 
``Nederlandse Organisatie voor Wetenschappelijk Onderzoek'' (NWO), 
is acknowledged. K. Saitoh is funded by the NWO-STW VICI grant 10828.

\end{acknowledgments}

\appendix
\section{Maximum attractive force}\label{sec:fmin_press}

The extreme loading and unloading branches are reflected by the outer triangle in Fig. \ref{fig:fnonadh}.
 Starting from a realized maximum overlap during loading, $\delta_{\rm max} < \delta^{p}_{\rm max}$, the unloading
 happens within the triangle, as can be characterized by a branch with stiffness 
\begin{equation}\label{eq:k2_del}
 k_2=k_1 + (k_{p}-k_1)  {\delta_{\rm max} } / { \delta^{p}_{\rm max} }
\end{equation}
 (as given in \cite{asingh13}). The elastic, reversible force along this branch is given by $k_2 (\delta-\delta_0)$ \cite{Luding08gm,asingh14}.
 The intermediate stiffness $k_2$ follows from a linear interpolation between $k_1$ and $k_p$, as explained in \cite{Luding08gm,asingh14}. The corresponding
 maximal attractive force is $f_{\rm m}= -k_c \delta_{\rm m}= -k_c \frac{({k_2-k_1})}{({k_2+k_c})} \delta_{\rm max} $.
 If we assume that the maximal overlap $\delta^{p}_{\rm max}$ is realized under a given
 external (compressive) pressure $p_{\rm max}$, then we can infer $\frac{p}{p_{\rm max}}=\frac{\delta_{\rm max}}{\delta^{p}_{\rm max}}$, with
 pressure $p$ being $p=k_1 \delta_{\rm max} /A$,\ $A$ being a representative area. This leads to realized maximal attractive force being
\begin{equation}\label{eq:fmin}
 f_{\rm m}= -k_c  \frac{({k_2-k_1})}{({k_2+k_c})} \frac{p}{p_{\rm max}} { \delta^{p}_{\rm max} }
\end{equation}
Using Eq.\ \eqref{eq:k2_del} in Eq.\ \eqref{eq:fmin}, we get
\begin{equation}\label{eq:fmin_p}
 f_{\rm m}= -k_c \frac{(k_p-k_1)\frac{p_{\rm max}}{k_1} \left(\frac{p}{p_{\rm max}} \right)^2 }{k_c+k_1+(k_p-k_1)\frac{p}{p_{\rm max}}}.
\end{equation}
This definition can be used to define a local Bond number as $Bo^a_{l}(p)=f_{\rm m}(p)/\langle f(p) \rangle$, where
 mean force at that pressure is discussed in Sec. \ref{sub:param}. This Bond number would be compared with various
 other definitions in Sec.  \ref{sub:param}.

\section{Cohesive force magnitude}\label{sec:fmin_magnitude}
In order to get a feeling for the magnitude of the adhesion forces 
in experimental systems, we resort to Ref.\ \cite{tahmasebpoora13}
and estimate the attractive force as:
$$F_{\rm vdW}=\frac{H d}{24 l^2} \approx 
  1.7 \times 10^{-10}\,\mathrm{N} \mathrm{ ~~or~~ } 1.7 \times 10^{-9}\,\mathrm{N}
       \, ,
$$
for SiO$_2$ particles with Hamaker constant $H=6.6 \times 10^{-20}$\,J, 
minimal inter-particle distance $l = s d \approx 4 \times 10^{-4} d$\, 
(order of surface roughness for 9 $\mu$m particles
of high quality is actually a factor ten smaller \cite{fuchs2014},
while realistic roughness can be even larger - the value of $s$
is just a rough estimate), and diameter $d=100$\,$\mu$m or $10$\,$\mu$m. 
Due to this assumed relative magnitude of surface roughness,
the adhesive force magnitude increases linearly with the (primary) particle 
diameter, while the gravitational force on the same particle
$$F_g = m g \approx
            10^{-8}\,\mathrm{N} \mathrm{ ~~or~~ } 10^{-11}\,\mathrm{N} 
    $$
decreases with the third power of the diameter, i.e., much faster.
So while 0.1 millimeter-size particles are dominated by gravity,
smaller 10 micron-size particles are dominated by their adhesion forces, as
reflected by the respective (single-particle) gravity Bond numbers 
$$Bo_g = F_{\rm vdW} / F_g
       = \frac{H}{4 \pi \rho s^2 d^4}
       = 1.7 \times 10^{-2}\, \mathrm{ ~~or~~ } 1.7 \times 10^{2}\,
       \, .
$$.

%

 \bibliography{./submit_20140417}

\begin{thebibliography}{10}%
\makeatletter
\providecommand \@ifxundefined [1]{%
 \ifx #1\undefined \expandafter \@firstoftwo
 \else \expandafter \@secondoftwo
\fi
}%
\providecommand \@ifnum [1]{%
 \ifnum #1\expandafter \@firstoftwo
 \else \expandafter \@secondoftwo
\fi
}%
\providecommand \enquote [1]{``#1''}%
\providecommand \bibnamefont  [1]{#1}%
\providecommand \bibfnamefont [1]{#1}%
\providecommand \citenamefont [1]{#1}%
\providecommand\href[0]{\@sanitize\@href}%
\providecommand\@href[1]{\endgroup\@@startlink{#1}\endgroup\@@href}%
\providecommand\@@href[1]{#1\@@endlink}%
\providecommand \@sanitize [0]{\begingroup\catcode`\&12\catcode`\#12\relax}%
\@ifxundefined \pdfoutput {\@firstoftwo}{%
 \@ifnum{\z@=\pdfoutput}{\@firstoftwo}{\@secondoftwo}%
}{%
 \providecommand\@@startlink[1]{\leavevmode\special{html:<a href="#1">}}%
 \providecommand\@@endlink[0]{\special{html:</a>}}%
}{%
 \providecommand\@@startlink[1]{%
  \leavevmode
  \pdfstartlink
   attr{/Border[0 0 1 ]/H/I/C[0 1 1]}%
   user{/Subtype/Link/A<</Type/Action/S/URI/URI(#1)>>}%
  \relax
 }%
 \providecommand\@@endlink[0]{\pdfendlink}%
}%
\providecommand \url  [0]{\begingroup\@sanitize \@url }%
\providecommand \@url [1]{\endgroup\@href {#1}{\urlprefix}}%
\providecommand \urlprefix [0]{URL }%
\providecommand \Eprint[0]{\href }%
\@ifxundefined \urlstyle {%
  \providecommand \doi [1]{doi:\discretionary{}{}{}#1}%
}{%
  \providecommand \doi [0]{doi:\discretionary{}{}{}\begingroup
  \urlstyle{rm}\Url }%
}%
\providecommand \doibase [0]{http://dx.doi.org/}%
\providecommand \Doi[1]{\href{\doibase#1}}%
\providecommand \bibAnnote [3]{%
  \BibitemShut{#1}%
  \begin{quotation}\noindent
    \textsc{Key:}\ #2\\\textsc{Annotation:}\ #3%
  \end{quotation}%
}%
\providecommand \bibAnnoteFile [2]{%
  \IfFileExists{#2}{\bibAnnote {#1} {#2} {\input{#2}}}{}%
}%
\providecommand \typeout [0]{\immediate \write \m@ne }%
\providecommand \selectlanguage [0]{\@gobble}%
\providecommand \bibinfo [0]{\@secondoftwo}%
\providecommand \bibfield [0]{\@secondoftwo}%
\providecommand \translation [1]{[#1]}%
\providecommand \BibitemOpen[0]{}%
\providecommand \bibitemStop [0]{}%
\providecommand \bibitemNoStop [0]{.\EOS\space}%
\providecommand \EOS [0]{\spacefactor3000\relax}%
\providecommand \BibitemShut [1]{\csname bibitem#1\endcsname}%
\bibitem{bridgwater80}%
  \BibitemOpen
  \bibfield{author}{%
  \bibinfo {author} {\bibfnamefont{J.}~\bibnamefont{Bridgwater}},\ }%
  \bibfield{journal}{%
  \bibinfo {journal} {G\'eotechnique}\ }%
  \textbf{\bibinfo {volume} {30}},\ \bibinfo {pages} {533} (\bibinfo {year}
  {1980})%
  \bibAnnoteFile{NoStop}{bridgwater80}%
\bibitem{howell99}%
  \BibitemOpen
  \bibfield{author}{%
  \bibinfo {author} {\bibfnamefont{D.}~\bibnamefont{Howell}}, \bibinfo {author}
  {\bibfnamefont{R.~P.}\ \bibnamefont{Behringer}},\ and\ \bibinfo {author}
  {\bibfnamefont{C.}~\bibnamefont{Veje}},\ }%
  \bibfield{journal}{%
  \bibinfo {journal} {Phys. Rev. Lett.}\ }%
  \textbf{\bibinfo {volume} {82}},\ \bibinfo {pages} {5241} (\bibinfo {year}
  {1999})%
  \bibAnnoteFile{NoStop}{howell99}%
\bibitem{schall2010shear}%
  \BibitemOpen
  \bibfield{author}{%
  \bibinfo {author} {\bibfnamefont{P.}~\bibnamefont{Schall}}\ and\ \bibinfo
  {author} {\bibfnamefont{M.}~\bibnamefont{van Hecke}},\ }%
  \bibfield{journal}{%
  \bibinfo {journal} {Ann. Rev. Fluid Mech.}\ }%
  \textbf{\bibinfo {volume} {42}} (\bibinfo {year} {2010})%
  \bibAnnoteFile{NoStop}{schall2010shear}%
\bibitem{Katgert08}%
  \BibitemOpen
  \bibfield{author}{%
  \bibinfo {author} {\bibfnamefont{G.}~\bibnamefont{Katgert}}, \bibinfo
  {author} {\bibfnamefont{M.~E.}\ \bibnamefont{M\"obius}},\ and\ \bibinfo
  {author} {\bibfnamefont{M.}~\bibnamefont{van Hecke}},\ }%
  \bibfield{journal}{%
  \bibinfo {journal} {Phys.\ Rev.\ Lett.}\ }%
  \textbf{\bibinfo {volume} {101}},\ \bibinfo {pages} {058301} (\bibinfo {year}
  {2008})%
  \bibAnnoteFile{NoStop}{Katgert08}%
\bibitem{Addad05}%
  \BibitemOpen
  \bibfield{author}{%
  \bibinfo {author} {\bibfnamefont{R.}~\bibnamefont{H\"ohler}}\ and\ \bibinfo
  {author} {\bibfnamefont{S.}~\bibnamefont{Cohen-Addad}},\ }%
  \bibfield{journal}{%
  \bibinfo {journal} {Journal of Physics: Condensed Matter}\ }%
  \textbf{\bibinfo {volume} {17}},\ \bibinfo {pages} {R1041} (\bibinfo {year}
  {2005})%
  \bibAnnoteFile{NoStop}{Addad05}%
\bibitem{Ovarlez09}%
  \BibitemOpen
  \bibfield{author}{%
  \bibinfo {author} {\bibfnamefont{G.}~\bibnamefont{Ovarlez}}, \bibinfo
  {author} {\bibfnamefont{S.}~\bibnamefont{Rodts}}, \bibinfo {author}
  {\bibfnamefont{X.}~\bibnamefont{Chateau}},\ and\ \bibinfo {author}
  {\bibfnamefont{P.}~\bibnamefont{Coussot}},\ }%
  \bibfield{journal}{%
  \bibinfo {journal} {Rheologica Acta}\ }%
  \textbf{\bibinfo {volume} {48}},\ \bibinfo {pages} {831} (\bibinfo {year}
  {2009}),\ ISSN \bibinfo {issn} {0035-4511}%
  \bibAnnoteFile{NoStop}{Ovarlez09}%
\bibitem{Dhont03}%
  \BibitemOpen
  \bibfield{author}{%
  \bibinfo {author} {\bibfnamefont{J.~K.~G.}\ \bibnamefont{Dhont}}, \bibinfo
  {author} {\bibfnamefont{M.~P.}\ \bibnamefont{Lettinga}}, \bibinfo {author}
  {\bibfnamefont{Z.}~\bibnamefont{Dogic}}, \bibinfo {author}
  {\bibfnamefont{T.~A.~J.}\ \bibnamefont{Lenstra}}, \bibinfo {author}
  {\bibfnamefont{H.}~\bibnamefont{Wang}}, \bibinfo {author}
  {\bibfnamefont{S.}~\bibnamefont{Rathgeber}}, \bibinfo {author}
  {\bibfnamefont{P.}~\bibnamefont{Carletto}}, \bibinfo {author}
  {\bibfnamefont{L.}~\bibnamefont{Willner}}, \bibinfo {author}
  {\bibfnamefont{H.}~\bibnamefont{Frielinghaus}},\ and\ \bibinfo {author}
  {\bibfnamefont{P.}~\bibnamefont{Lindner}},\ }%
  \bibfield{journal}{%
  \bibinfo {journal} {Faraday Discuss.}\ }%
  \textbf{\bibinfo {volume} {123}},\ \bibinfo {pages} {157} (\bibinfo {year}
  {2003})%
  \bibAnnoteFile{NoStop}{Dhont03}%
\bibitem{Mandl77}%
  \BibitemOpen
  \bibfield{author}{%
  \bibinfo {author} {\bibfnamefont{G.}~\bibnamefont{Mandl}}, \bibinfo {author}
  {\bibfnamefont{L.}~\bibnamefont{Jong}},\ and\ \bibinfo {author}
  {\bibfnamefont{A.}~\bibnamefont{Maltha}},\ }%
  \bibfield{journal}{%
  \bibinfo {journal} {Rock Mech.}\ }%
  \textbf{\bibinfo {volume} {9}},\ \bibinfo {pages} {95} (\bibinfo {year}
  {1977}),\ ISSN \bibinfo {issn} {0035-7448}%
  \bibAnnoteFile{NoStop}{Mandl77}%
\bibitem{bardet91}%
  \BibitemOpen
  \bibfield{author}{%
  \bibinfo {author} {\bibfnamefont{J.~P.}\ \bibnamefont{Bardet}}\ and\ \bibinfo
  {author} {\bibfnamefont{J.}~\bibnamefont{Proubet}},\ }%
  \bibfield{journal}{%
  \bibinfo {journal} {G\'eotechnique}\ }%
  \textbf{\bibinfo {volume} {41}},\ \bibinfo {pages} {599} (\bibinfo {year}
  {1991})%
  \bibAnnoteFile{NoStop}{bardet91}%
\bibitem{oda98}%
  \BibitemOpen
  \bibfield{author}{%
  \bibinfo {author} {\bibfnamefont{M.}~\bibnamefont{Oda}}\ and\ \bibinfo
  {author} {\bibfnamefont{H.}~\bibnamefont{Kazama}},\ }%
  \bibfield{journal}{%
  \bibinfo {journal} {G\'eotechnique}\ }%
  \textbf{\bibinfo {volume} {48}},\ \bibinfo {pages} {465} (\bibinfo {year}
  {1998})%
  \bibAnnoteFile{NoStop}{oda98}%
\bibitem{Scott96}%
  \BibitemOpen
  \bibfield{author}{%
  \bibinfo {author} {\bibfnamefont{D.~R.}\ \bibnamefont{Scott}},\ }%
  \bibfield{journal}{%
  \bibinfo {journal} {Nature}\ }%
  \textbf{\bibinfo {volume} {381}},\ \bibinfo {pages} {592} (\bibinfo {year}
  {1996})%
  \bibAnnoteFile{NoStop}{Scott96}%
\bibitem{muhlhaus87b}%
  \BibitemOpen
  \bibfield{author}{%
  \bibinfo {author} {\bibfnamefont{H.-B.}\ \bibnamefont{M\"uhlhaus}}\ and\
  \bibinfo {author} {\bibfnamefont{I.}~\bibnamefont{Vardoulakis}},\ }%
  \bibfield{journal}{%
  \bibinfo {journal} {G\'{e}otechnique}\ }%
  \textbf{\bibinfo {volume} {3}},\ \bibinfo {pages} {271} (\bibinfo {year}
  {1987})%
  \bibAnnoteFile{NoStop}{muhlhaus87b}%
\bibitem{latzel00}%
  \BibitemOpen
  \bibfield{author}{%
  \bibinfo {author} {\bibfnamefont{M.}~\bibnamefont{L\"atzel}}, \bibinfo
  {author} {\bibfnamefont{S.}~\bibnamefont{Luding}},\ and\ \bibinfo {author}
  {\bibfnamefont{H.~J.}\ \bibnamefont{Herrmann}},\ }%
  \bibfield{journal}{%
  \bibinfo {journal} {Granular Matter}\ }%
  \textbf{\bibinfo {volume} {2}},\ \bibinfo {pages} {123} (\bibinfo {year}
  {2000})%
  \bibAnnoteFile{NoStop}{latzel00}%
\bibitem{latzel03}%
  \BibitemOpen
  \bibfield{author}{%
  \bibinfo {author} {\bibfnamefont{M.}~\bibnamefont{L\"atzel}}, \bibinfo
  {author} {\bibfnamefont{S.}~\bibnamefont{Luding}}, \bibinfo {author}
  {\bibfnamefont{H.~J.}\ \bibnamefont{Herrmann}}, \bibinfo {author}
  {\bibfnamefont{D.~W.}\ \bibnamefont{Howell}},\ and\ \bibinfo {author}
  {\bibfnamefont{R.~P.}\ \bibnamefont{Behringer}},\ }%
  \bibfield{journal}{%
  \bibinfo {journal} {Eur. Phys. J. E}\ }%
  \textbf{\bibinfo {volume} {11}},\ \bibinfo {pages} {325} (\bibinfo {year}
  {2003})%
  \bibAnnoteFile{NoStop}{latzel03}%
\bibitem{fenistein03}%
  \BibitemOpen
  \bibfield{author}{%
  \bibinfo {author} {\bibfnamefont{D.}~\bibnamefont{Fenistein}}\ and\ \bibinfo
  {author} {\bibfnamefont{M.}~\bibnamefont{van Hecke}},\ }%
  \bibfield{journal}{%
  \bibinfo {journal} {Nature}\ }%
  \textbf{\bibinfo {volume} {425}},\ \bibinfo {pages} {256} (\bibinfo {year}
  {2003})%
  \bibAnnoteFile{NoStop}{fenistein03}%
\bibitem{mcnamara1996dynamics}%
  \BibitemOpen
  \bibfield{author}{%
  \bibinfo {author} {\bibfnamefont{S.}~\bibnamefont{McNamara}}\ and\ \bibinfo
  {author} {\bibfnamefont{W.~R.}\ \bibnamefont{Young}},\ }%
  \bibfield{journal}{%
  \bibinfo {journal} {Phys. Rev. E}\ }%
  \textbf{\bibinfo {volume} {53}},\ \bibinfo {pages} {5089} (\bibinfo {year}
  {1996})%
  \bibAnnoteFile{NoStop}{mcnamara1996dynamics}%
\bibitem{mcnamara1994inelastic}%
  \BibitemOpen
  \bibfield{author}{%
  \bibinfo {author} {\bibfnamefont{S.}~\bibnamefont{McNamara}}\ and\ \bibinfo
  {author} {\bibfnamefont{W.~R.}\ \bibnamefont{Young}},\ }%
  \bibfield{journal}{%
  \bibinfo {journal} {Phys. Rev. E}\ }%
  \textbf{\bibinfo {volume} {50}},\ \bibinfo {pages} {R28} (\bibinfo {year}
  {1994})%
  \bibAnnoteFile{NoStop}{mcnamara1994inelastic}%
\bibitem{KamrinKoval2012}%
  \BibitemOpen
  \bibfield{author}{%
  \bibinfo {author} {\bibfnamefont{K.}~\bibnamefont{Kamrin}}\ and\ \bibinfo
  {author} {\bibfnamefont{G.}~\bibnamefont{Koval}},\ }%
  \bibfield{journal}{%
  \bibinfo {journal} {Phys. Rev. Lett.}\ }%
  \textbf{\bibinfo {volume} {108}},\ \bibinfo {pages} {178301} (\bibinfo {year}
  {2012})%
  \bibAnnoteFile{NoStop}{KamrinKoval2012}%
\bibitem{kamrin2013predictive}%
  \BibitemOpen
  \bibfield{author}{%
  \bibinfo {author} {\bibfnamefont{D.~L.}\ \bibnamefont{Henann}}\ and\ \bibinfo
  {author} {\bibfnamefont{K.}~\bibnamefont{Kamrin}},\ }%
  \bibfield{journal}{%
  \bibinfo {journal} {P. N. A. S.}}%
   (\bibinfo {year} {2013})%
  \bibAnnoteFile{NoStop}{kamrin2013predictive}%
\bibitem{Quintanilla03}%
  \BibitemOpen
  \bibfield{author}{%
  \bibinfo {author} {\bibfnamefont{M.~A.~S.}\ \bibnamefont{Quintanilla}},
  \bibinfo {author} {\bibfnamefont{A.}~\bibnamefont{Castellanos}},\ and\
  \bibinfo {author} {\bibfnamefont{J.~M.}\ \bibnamefont{Valverde}},\ }%
  \bibfield{journal}{%
  \bibinfo {journal} {PAMM}\ }%
  \textbf{\bibinfo {volume} {3}},\ \bibinfo {pages} {206} (\bibinfo {year}
  {2003})%
  \bibAnnoteFile{NoStop}{Quintanilla03}%
\bibitem{Valverdejammingpowder04}%
  \BibitemOpen
  \bibfield{author}{%
  \bibinfo {author} {\bibfnamefont{J.~M.}\ \bibnamefont{Valverde}}, \bibinfo
  {author} {\bibfnamefont{M.~A.~S.}\ \bibnamefont{Quintanilla}},\ and\ \bibinfo
  {author} {\bibfnamefont{A.}~\bibnamefont{Castellanos}},\ }%
  \bibfield{journal}{%
  \bibinfo {journal} {Phys. Rev. Lett.}\ }%
  \textbf{\bibinfo {volume} {92}},\ \bibinfo {pages} {258303} (\bibinfo {year}
  {2004})%
  \bibAnnoteFile{NoStop}{Valverdejammingpowder04}%
\bibitem{Castellanos05}%
  \BibitemOpen
  \bibfield{author}{%
  \bibinfo {author} {\bibfnamefont{A.}~\bibnamefont{Castellanos}},\ }%
  \bibfield{journal}{%
  \bibinfo {journal} {Advances in Physics}\ }%
  \textbf{\bibinfo {volume} {54}},\ \bibinfo {pages} {263} (\bibinfo {year}
  {2005})%
  \bibAnnoteFile{NoStop}{Castellanos05}%
\bibitem{herminghaus05}%
  \BibitemOpen
  \bibfield{author}{%
  \bibinfo {author} {\bibfnamefont{S.}~\bibnamefont{Herminghaus}},\ }%
  \bibfield{journal}{%
  \bibinfo {journal} {Advances in Physics}\ }%
  \textbf{\bibinfo {volume} {54}},\ \bibinfo {pages} {221} (\bibinfo {year}
  {2005})%
  \bibAnnoteFile{NoStop}{herminghaus05}%
\bibitem{Micela05}%
  \BibitemOpen
  \bibfield{author}{%
  \bibinfo {author} {\bibfnamefont{C.}~\bibnamefont{Miclea}}, \bibinfo {author}
  {\bibfnamefont{C.}~\bibnamefont{Tanasoiu}}, \bibinfo {author}
  {\bibfnamefont{C.}~\bibnamefont{Miclea}}, \bibinfo {author}
  {\bibfnamefont{F.}~\bibnamefont{Sima}}, \bibinfo {author}
  {\bibfnamefont{M.}~\bibnamefont{Cioangher}},\ and\ \bibinfo {author}
  {\bibfnamefont{A.}~\bibnamefont{Gheorghiu}},\ }%
  in\ \emph{\bibinfo {booktitle} {Powders and Grains 2005}}\ (\bibinfo
  {publisher} {Balkema},\ \bibinfo {year} {2005})%
  \bibAnnoteFile{NoStop}{Micela05}%
\bibitem{brendel-2011}%
  \BibitemOpen
  \bibfield{author}{%
  \bibinfo {author} {\bibfnamefont{L.}~\bibnamefont{Brendel}}, \bibinfo
  {author} {\bibfnamefont{J.}~\bibnamefont{T\"{o}r\"{o}k}}, \bibinfo {author}
  {\bibfnamefont{R.}~\bibnamefont{Kirsch}},\ and\ \bibinfo {author}
  {\bibfnamefont{U.}~\bibnamefont{Br\"ockel}},\ }%
  \bibfield{journal}{%
  \bibinfo {journal} {Granular Matter}\ }%
  \textbf{\bibinfo {volume} {13}},\ \bibinfo {pages} {777} (\bibinfo {year}
  {2011}),\ ISSN \bibinfo {issn} {1434-5021}%
  \bibAnnoteFile{NoStop}{brendel-2011}%
\bibitem{hatzes91}%
  \BibitemOpen
  \bibfield{author}{%
  \bibinfo {author} {\bibfnamefont{A.}~\bibnamefont{Hatzes}}, \bibinfo {author}
  {\bibfnamefont{F.}~\bibnamefont{Bridges}}, \bibinfo {author}
  {\bibfnamefont{D.}~\bibnamefont{Lin}},\ and\ \bibinfo {author}
  {\bibfnamefont{S.}~\bibnamefont{Sachtjen}},\ }%
  \bibfield{journal}{%
  \bibinfo {journal} {Icarus}\ }%
  \textbf{\bibinfo {volume} {89}},\ \bibinfo {pages} {113} (\bibinfo {year}
  {1991})%
  \bibAnnoteFile{NoStop}{hatzes91}%
\bibitem{Spaepen1977}%
  \BibitemOpen
  \bibfield{author}{%
  \bibinfo {author} {\bibfnamefont{F.}~\bibnamefont{Spaepen}},\ }%
  \bibfield{journal}{%
  \bibinfo {journal} {Acta Meta.}\ }%
  \textbf{\bibinfo {volume} {25}},\ \bibinfo {pages} {407} (\bibinfo {year}
  {1977})%
  \bibAnnoteFile{NoStop}{Spaepen1977}%
\bibitem{LiSBglasses}%
  \BibitemOpen
  \bibfield{author}{%
  \bibinfo {author} {\bibfnamefont{J.}~\bibnamefont{Li}}, \bibinfo {author}
  {\bibfnamefont{F.}~\bibnamefont{Spaepen}},\ and\ \bibinfo {author}
  {\bibfnamefont{T.~C.}\ \bibnamefont{Hufnagel}},\ }%
  \bibfield{journal}{%
  \bibinfo {journal} {Philos. Mag. A}\ }%
  \textbf{\bibinfo {volume} {82}},\ \bibinfo {pages} {2623} (\bibinfo {year}
  {2002})%
  \bibAnnoteFile{NoStop}{LiSBglasses}%
\bibitem{Becu06}%
  \BibitemOpen
  \bibfield{author}{%
  \bibinfo {author} {\bibfnamefont{L.}~\bibnamefont{B\'{e}cu}}, \bibinfo
  {author} {\bibfnamefont{S.}~\bibnamefont{Manneville}},\ and\ \bibinfo
  {author} {\bibfnamefont{A.}~\bibnamefont{Colin}},\ }%
  \bibfield{journal}{%
  \bibinfo {journal} {Phys. Rev. Lett.}\ }%
  \textbf{\bibinfo {volume} {96}},\ \bibinfo {pages} {138302} (\bibinfo {year}
  {2006})%
  \bibAnnoteFile{NoStop}{Becu06}%
\bibitem{Chaudhuri2012}%
  \BibitemOpen
  \bibfield{author}{%
  \bibinfo {author} {\bibfnamefont{P.}~\bibnamefont{Chaudhuri}}, \bibinfo
  {author} {\bibfnamefont{L.}~\bibnamefont{Berthier}},\ and\ \bibinfo {author}
  {\bibfnamefont{L.}~\bibnamefont{Bocquet}},\ }%
  \bibfield{journal}{%
  \bibinfo {journal} {Phys. Rev. E}\ }%
  \textbf{\bibinfo {volume} {85}},\ \bibinfo {pages} {021503} (\bibinfo {year}
  {2012})%
  \bibAnnoteFile{NoStop}{Chaudhuri2012}%
\bibitem{vermant2001large}%
  \BibitemOpen
  \bibfield{author}{%
  \bibinfo {author} {\bibfnamefont{J.}~\bibnamefont{Vermant}},\ }%
  \bibfield{journal}{%
  \bibinfo {journal} {Current opinion in colloid \& interface science}\ }%
  \textbf{\bibinfo {volume} {6}},\ \bibinfo {pages} {489} (\bibinfo {year}
  {2001})%
  \bibAnnoteFile{NoStop}{vermant2001large}%
\bibitem{hohler2005rheology}%
  \BibitemOpen
  \bibfield{author}{%
  \bibinfo {author} {\bibfnamefont{R.}~\bibnamefont{H{\"o}hler}}\ and\ \bibinfo
  {author} {\bibfnamefont{S.}~\bibnamefont{Cohen-Addad}},\ }%
  \bibfield{journal}{%
  \bibinfo {journal} {Journal of Physics: Condensed Matter}\ }%
  \textbf{\bibinfo {volume} {17}},\ \bibinfo {pages} {R1041} (\bibinfo {year}
  {2005})%
  \bibAnnoteFile{NoStop}{hohler2005rheology}%
\bibitem{coussot2010physical}%
  \BibitemOpen
  \bibfield{author}{%
  \bibinfo {author} {\bibfnamefont{P.}~\bibnamefont{Coussot}}\ and\ \bibinfo
  {author} {\bibfnamefont{G.}~\bibnamefont{Ovarlez}},\ }%
  \bibfield{journal}{%
  \bibinfo {journal} {The European Physical Journal E}\ }%
  \textbf{\bibinfo {volume} {33}},\ \bibinfo {pages} {183} (\bibinfo {year}
  {2010})%
  \bibAnnoteFile{NoStop}{coussot2010physical}%
\bibitem{Nicolas10}%
  \BibitemOpen
  \bibfield{author}{%
  \bibinfo {author} {\bibfnamefont{N.}~\bibnamefont{Estrada}}, \bibinfo
  {author} {\bibfnamefont{A.}~\bibnamefont{Lizcano}},\ and\ \bibinfo {author}
  {\bibfnamefont{A.}~\bibnamefont{Taboada}},\ }%
  \bibfield{journal}{%
  \bibinfo {journal} {Phys. Rev. E}\ }%
  \textbf{\bibinfo {volume} {82}},\ \bibinfo {pages} {011303} (\bibinfo {year}
  {2010})%
  \bibAnnoteFile{NoStop}{Nicolas10}%
\bibitem{Roman12}%
  \BibitemOpen
  \bibfield{author}{%
  \bibinfo {author} {\bibfnamefont{R.}~\bibnamefont{Mani}}, \bibinfo {author}
  {\bibfnamefont{D.}~\bibnamefont{Kadau}}, \bibinfo {author}
  {\bibfnamefont{D.}~\bibnamefont{Or}},\ and\ \bibinfo {author}
  {\bibfnamefont{H.~J.}\ \bibnamefont{Herrmann}},\ }%
  \bibfield{journal}{%
  \bibinfo {journal} {Phys. Rev. Lett.}\ }%
  \textbf{\bibinfo {volume} {109}},\ \bibinfo {pages} {248001} (\bibinfo {year}
  {2012})%
  \bibAnnoteFile{NoStop}{Roman12}%
\bibitem{Fabian13}%
  \BibitemOpen
  \bibfield{author}{%
  \bibinfo {author} {\bibfnamefont{R.}~\bibnamefont{Schwarze}}, \bibinfo
  {author} {\bibfnamefont{A.}~\bibnamefont{Gladkyy}}, \bibinfo {author}
  {\bibfnamefont{F.}~\bibnamefont{Uhlig}},\ and\ \bibinfo {author}
  {\bibfnamefont{S.}~\bibnamefont{Luding}},\ }%
  \bibfield{journal}{%
  \bibinfo {journal} {Granular Matter}\ }%
  \textbf{\bibinfo {volume} {15}} (\bibinfo {year} {2013}),\ ISSN \bibinfo
  {issn} {1434-5021}%
  \bibAnnoteFile{NoStop}{Fabian13}%
\bibitem{Yuan13}%
  \BibitemOpen
  \bibfield{author}{%
  \bibinfo {author} {\bibfnamefont{J.}~\bibnamefont{Yuan}}, \bibinfo {author}
  {\bibfnamefont{Q.}~\bibnamefont{Zhang}}, \bibinfo {author}
  {\bibfnamefont{B.}~\bibnamefont{Li}},\ and\ \bibinfo {author}
  {\bibfnamefont{X.}~\bibnamefont{Zhao}},\ }%
  \bibfield{journal}{%
  \bibinfo {journal} {Bulletin of Engineering Geology and the Environment}\ }%
  \textbf{\bibinfo {volume} {72}},\ \bibinfo {pages} {107} (\bibinfo {year}
  {2013}),\ ISSN \bibinfo {issn} {1435-9529}%
  \bibAnnoteFile{NoStop}{Yuan13}%
\bibitem{liu1995force}%
  \BibitemOpen
  \bibfield{author}{%
  \bibinfo {author} {\bibfnamefont{C.~H.}\ \bibnamefont{Liu}}, \bibinfo
  {author} {\bibfnamefont{S.~R.}\ \bibnamefont{Nagel}}, \bibinfo {author}
  {\bibfnamefont{D.~A.}\ \bibnamefont{Schecter}}, \bibinfo {author}
  {\bibfnamefont{S.~N.}\ \bibnamefont{Coppersmith}}, \bibinfo {author}
  {\bibfnamefont{S.}~\bibnamefont{Majumdar}}, \bibinfo {author}
  {\bibfnamefont{O.}~\bibnamefont{Narayan}},\ and\ \bibinfo {author}
  {\bibfnamefont{T.~A.}\ \bibnamefont{Witten}},\ }%
  \bibfield{journal}{%
  \bibinfo {journal} {Science}\ }%
  \textbf{\bibinfo {volume} {269}},\ \bibinfo {pages} {513} (\bibinfo {year}
  {1995})%
  \bibAnnoteFile{NoStop}{liu1995force}%
\bibitem{mueth98}%
  \BibitemOpen
  \bibfield{author}{%
  \bibinfo {author} {\bibfnamefont{D.~M.}\ \bibnamefont{Mueth}}, \bibinfo
  {author} {\bibfnamefont{H.~M.}\ \bibnamefont{Jaeger}},\ and\ \bibinfo
  {author} {\bibfnamefont{S.~R.}\ \bibnamefont{Nagel}},\ }%
  \bibfield{journal}{%
  \bibinfo {journal} {Phys. Rev. E.}\ }%
  \textbf{\bibinfo {volume} {57}},\ \bibinfo {pages} {3164} (\bibinfo {year}
  {1998})%
  \bibAnnoteFile{NoStop}{mueth98}%
\bibitem{lovoll0000force}%
  \BibitemOpen
  \bibfield{author}{%
  \bibinfo {author} {\bibfnamefont{G.}~\bibnamefont{L\o{}voll}}, \bibinfo
  {author} {\bibfnamefont{K.~J.}\ \bibnamefont{M\aa{}l\o{}y}},\ and\ \bibinfo
  {author} {\bibfnamefont{E.~G.}\ \bibnamefont{Flekk\o{}y}},\ }%
  \bibfield{journal}{%
  \bibinfo {journal} {Phys. Rev. E}\ }%
  \textbf{\bibinfo {volume} {60}},\ \bibinfo {pages} {5872} (\bibinfo {year}
  {1999})%
  \bibAnnoteFile{NoStop}{lovoll0000force}%
\bibitem{blair01b}%
  \BibitemOpen
  \bibfield{author}{%
  \bibinfo {author} {\bibfnamefont{D.~L.}\ \bibnamefont{Blair}}, \bibinfo
  {author} {\bibfnamefont{N.~W.}\ \bibnamefont{Mueggenburg}}, \bibinfo {author}
  {\bibfnamefont{A.~H.}\ \bibnamefont{Marshall}}, \bibinfo {author}
  {\bibfnamefont{H.~M.}\ \bibnamefont{Jaeger}},\ and\ \bibinfo {author}
  {\bibfnamefont{S.~R.}\ \bibnamefont{Nagel}},\ }%
  \bibfield{journal}{%
  \bibinfo {journal} {Phys. Rev. E}\ }%
  \textbf{\bibinfo {volume} {63}},\ \bibinfo {pages} {041304} (\bibinfo {year}
  {2001}),\ \bibinfo {note} {cond-mat/0009313}%
  \bibAnnoteFile{NoStop}{blair01b}%
\bibitem{majmudar2005contact}%
  \BibitemOpen
  \bibfield{author}{%
  \bibinfo {author} {\bibfnamefont{T.~S.}\ \bibnamefont{Majmudar}}\ and\
  \bibinfo {author} {\bibfnamefont{R.~P.}\ \bibnamefont{Behringer}},\ }%
  \bibfield{journal}{%
  \bibinfo {journal} {Nature}\ }%
  \textbf{\bibinfo {volume} {435}},\ \bibinfo {pages} {1079} (\bibinfo {year}
  {2005})%
  \bibAnnoteFile{NoStop}{majmudar2005contact}%
\bibitem{liffman1992force}%
  \BibitemOpen
  \bibfield{author}{%
  \bibinfo {author} {\bibfnamefont{K.}~\bibnamefont{Liffman}}, \bibinfo
  {author} {\bibfnamefont{D.~Y.~C.}\ \bibnamefont{Chan}},\ and\ \bibinfo
  {author} {\bibfnamefont{B.~D.}\ \bibnamefont{Hughes}},\ }%
  \bibfield{journal}{%
  \bibinfo {journal} {Powder Technol.}\ }%
  \textbf{\bibinfo {volume} {72}},\ \bibinfo {pages} {255} (\bibinfo {year}
  {1992})%
  \bibAnnoteFile{NoStop}{liffman1992force}%
\bibitem{radjai98b}%
  \BibitemOpen
  \bibfield{author}{%
  \bibinfo {author} {\bibfnamefont{F.}~\bibnamefont{Radjai}}, \bibinfo {author}
  {\bibfnamefont{D.~E.}\ \bibnamefont{Wolf}}, \bibinfo {author}
  {\bibfnamefont{M.}~\bibnamefont{Jean}},\ and\ \bibinfo {author}
  {\bibfnamefont{J.-J.}\ \bibnamefont{Moreau}},\ }%
  \bibfield{journal}{%
  \bibinfo {journal} {Phys. Rev. Lett.}\ }%
  \textbf{\bibinfo {volume} {80}},\ \bibinfo {pages} {61} (\bibinfo {year}
  {1998})%
  \bibAnnoteFile{NoStop}{radjai98b}%
\bibitem{silbert2002statistics}%
  \BibitemOpen
  \bibfield{author}{%
  \bibinfo {author} {\bibfnamefont{L.~E.}\ \bibnamefont{Silbert}}, \bibinfo
  {author} {\bibfnamefont{G.~S.}\ \bibnamefont{Grest}},\ and\ \bibinfo {author}
  {\bibfnamefont{J.~W.}\ \bibnamefont{Landry}},\ }%
  \bibfield{journal}{%
  \bibinfo {journal} {Phys. Rev. E}\ }%
  \textbf{\bibinfo {volume} {66}},\ \bibinfo {pages} {061303} (\bibinfo {year}
  {2002})%
  \bibAnnoteFile{NoStop}{silbert2002statistics}%
\bibitem{Snoeijer03}%
  \BibitemOpen
  \bibfield{author}{%
  \bibinfo {author} {\bibfnamefont{J.~H.}\ \bibnamefont{Snoeijer}}, \bibinfo
  {author} {\bibfnamefont{M.}~\bibnamefont{van Hecke}}, \bibinfo {author}
  {\bibfnamefont{E.}~\bibnamefont{Somfai}},\ and\ \bibinfo {author}
  {\bibfnamefont{W.}~\bibnamefont{van Saarloos}},\ }%
  \bibfield{journal}{%
  \bibinfo {journal} {Phys. Rev. E}\ }%
  \textbf{\bibinfo {volume} {67}},\ \bibinfo {pages} {030302} (\bibinfo {year}
  {2003})%
  \bibAnnoteFile{NoStop}{Snoeijer03}%
\bibitem{Silber10}%
  \BibitemOpen
  \bibfield{author}{%
  \bibinfo {author} {\bibfnamefont{L.~E.}\ \bibnamefont{Silbert}},\ }%
  \bibfield{journal}{%
  \bibinfo {journal} {Soft Matter}\ }%
  \textbf{\bibinfo {volume} {6}},\ \bibinfo {pages} {2918} (\bibinfo {year}
  {2010})%
  \bibAnnoteFile{NoStop}{Silber10}%
\bibitem{trappe2001jamming}%
  \BibitemOpen
  \bibfield{author}{%
  \bibinfo {author} {\bibfnamefont{V.}~\bibnamefont{Trappe}}, \bibinfo {author}
  {\bibfnamefont{V.}~\bibnamefont{Prasad}}, \bibinfo {author}
  {\bibfnamefont{L.}~\bibnamefont{Cipelletti}}, \bibinfo {author}
  {\bibfnamefont{P.~N.}\ \bibnamefont{Segre}},\ and\ \bibinfo {author}
  {\bibfnamefont{D.~A.}\ \bibnamefont{Weitz}},\ }%
  \bibfield{journal}{%
  \bibinfo {journal} {Nature}\ }%
  \textbf{\bibinfo {volume} {411}},\ \bibinfo {pages} {772} (\bibinfo {year}
  {2001})%
  \bibAnnoteFile{NoStop}{trappe2001jamming}%
\bibitem{Radjaiwet06}%
  \BibitemOpen
  \bibfield{author}{%
  \bibinfo {author} {\bibfnamefont{V.}~\bibnamefont{Richefeu}}, \bibinfo
  {author} {\bibfnamefont{M.~S.~E.}\ \bibnamefont{Youssoufi}},\ and\ \bibinfo
  {author} {\bibfnamefont{F.}~\bibnamefont{Radja\"\i}},\ }%
  \bibfield{journal}{%
  \bibinfo {journal} {Phys. Rev. E}\ }%
  \textbf{\bibinfo {volume} {73}},\ \bibinfo {pages} {051304} (\bibinfo {year}
  {2006})%
  \bibAnnoteFile{NoStop}{Radjaiwet06}%
\bibitem{Rouxcoh07}%
  \BibitemOpen
  \bibfield{author}{%
  \bibinfo {author} {\bibfnamefont{F.~A.}\ \bibnamefont{Gilabert}}, \bibinfo
  {author} {\bibfnamefont{J.-N.}\ \bibnamefont{Roux}},\ and\ \bibinfo {author}
  {\bibfnamefont{A.}~\bibnamefont{Castellanos}},\ }%
  \bibfield{journal}{%
  \bibinfo {journal} {Phys. Rev. E}\ }%
  \textbf{\bibinfo {volume} {75}},\ \bibinfo {pages} {011303} (\bibinfo {year}
  {2007})%
  \bibAnnoteFile{NoStop}{Rouxcoh07}%
\bibitem{ABYucohfor08}%
  \BibitemOpen
  \bibfield{author}{%
  \bibinfo {author} {\bibfnamefont{R.~Y.}\ \bibnamefont{Yang}}, \bibinfo
  {author} {\bibfnamefont{R.~P.}\ \bibnamefont{Zou}}, \bibinfo {author}
  {\bibfnamefont{A.~B.}\ \bibnamefont{Yu}},\ and\ \bibinfo {author}
  {\bibfnamefont{S.~K.}\ \bibnamefont{Choi}},\ }%
  \bibfield{journal}{%
  \bibinfo {journal} {Phys. Rev. E}\ }%
  \textbf{\bibinfo {volume} {78}},\ \bibinfo {pages} {031302} (\bibinfo {year}
  {2008})%
  \bibAnnoteFile{NoStop}{ABYucohfor08}%
\bibitem{Radjaidry-wet10}%
  \BibitemOpen
  \bibfield{author}{%
  \bibinfo {author} {\bibfnamefont{F.}~\bibnamefont{Radjai}}, \bibinfo {author}
  {\bibfnamefont{V.}~\bibnamefont{Topin}}, \bibinfo {author}
  {\bibfnamefont{V.}~\bibnamefont{Richefeu}}, \bibinfo {author}
  {\bibfnamefont{C.}~\bibnamefont{Voivret}}, \bibinfo {author}
  {\bibfnamefont{J.}~\bibnamefont{Delenne}}, \bibinfo {author}
  {\bibfnamefont{E.}~\bibnamefont{Az\'ema}},\ and\ \bibinfo {author}
  {\bibfnamefont{S.}~\bibnamefont{El~Youssoufi}},\ }%
  \bibfield{journal}{%
  \bibinfo {journal} {AIP Conf. Proc.}\ }%
  \textbf{\bibinfo {volume} {1227}},\ \bibinfo {pages} {240} (\bibinfo {year}
  {2010})%
  \bibAnnoteFile{NoStop}{Radjaidry-wet10}%
\bibitem{luding11}%
  \BibitemOpen
  \bibfield{author}{%
  \bibinfo {author} {\bibfnamefont{S.}~\bibnamefont{Luding}}\ and\ \bibinfo
  {author} {\bibfnamefont{F.}~\bibnamefont{Alonso-Marroqu{\'\i}n}},\ }%
  \bibfield{journal}{%
  \bibinfo {journal} {Granular Matter}\ }%
  \textbf{\bibinfo {volume} {13}},\ \bibinfo {pages} {109} (\bibinfo {year}
  {2011})%
  \bibAnnoteFile{NoStop}{luding11}%
\bibitem{Nase01}%
  \BibitemOpen
  \bibfield{author}{%
  \bibinfo {author} {\bibfnamefont{S.~T.}\ \bibnamefont{Nase}}, \bibinfo
  {author} {\bibfnamefont{W.~L.}\ \bibnamefont{Vargas}}, \bibinfo {author}
  {\bibfnamefont{A.~A.}\ \bibnamefont{Abatan}},\ and\ \bibinfo {author}
  {\bibfnamefont{J.}~\bibnamefont{McCarthy}},\ }%
  \bibfield{journal}{%
  \bibinfo {journal} {Powder Technol.}\ }%
  \textbf{\bibinfo {volume} {116}},\ \bibinfo {pages} {214 } (\bibinfo {year}
  {2001})%
  \bibAnnoteFile{NoStop}{Nase01}%
\bibitem{Rognon08}%
  \BibitemOpen
  \bibfield{author}{%
  \bibinfo {author} {\bibfnamefont{P.~G.}\ \bibnamefont{Rognon}}, \bibinfo
  {author} {\bibfnamefont{J.}~\bibnamefont{Roux}},\ and\ \bibinfo {author}
  {\bibfnamefont{M.}~\bibnamefont{Naa\'im}},\ }%
  \bibfield{journal}{%
  \bibinfo {journal} {J. of Fluid Mech.}\ }%
  \textbf{\bibinfo {volume} {596}},\ \bibinfo {pages} {21 } (\bibinfo {year}
  {2008})%
  \bibAnnoteFile{NoStop}{Rognon08}%
\bibitem{allen87}%
  \BibitemOpen
  \bibfield{author}{%
  \bibinfo {author} {\bibfnamefont{M.~P.}\ \bibnamefont{Allen}}\ and\ \bibinfo
  {author} {\bibfnamefont{D.~J.}\ \bibnamefont{Tildesley}},\ }%
  \emph{\bibinfo {title} {Computer Simulation of Liquids}}\ (\bibinfo
  {publisher} {Oxford University Press},\ \bibinfo {year} {1987})%
  \bibAnnoteFile{NoStop}{allen87}%
\bibitem{cundall71}%
  \BibitemOpen
  \bibfield{author}{%
  \bibinfo {author} {\bibfnamefont{P.~A.}\ \bibnamefont{Cundall}},\ }%
  in\ \emph{\bibinfo {booktitle} {Proc. Symp. Int. Rock Mech.}},\ Vol.~\bibinfo
  {volume} {2}\ (\bibinfo {address} {Nancy},\ \bibinfo {year} {1971})%
  \bibAnnoteFile{NoStop}{cundall71}%
\bibitem{asingh13}%
  \BibitemOpen
  \bibfield{author}{%
  \bibinfo {author} {\bibfnamefont{A.}~\bibnamefont{Singh}}, \bibinfo {author}
  {\bibfnamefont{V.}~\bibnamefont{Magnanimo}},\ and\ \bibinfo {author}
  {\bibfnamefont{S.}~\bibnamefont{Luding}},\ }%
  \bibfield{journal}{%
  \bibinfo {journal} {AIP Conf. Proc.}\ }%
  \textbf{\bibinfo {volume} {1542}},\ \bibinfo {pages} {682} (\bibinfo {year}
  {2013})%
  \bibAnnoteFile{NoStop}{asingh13}%
\bibitem{Luding08gm}%
  \BibitemOpen
  \bibfield{author}{%
  \bibinfo {author} {\bibfnamefont{S.}~\bibnamefont{Luding}},\ }%
  \bibfield{journal}{%
  \bibinfo {journal} {Granular Matter}\ }%
  \textbf{\bibinfo {volume} {10}},\ \bibinfo {pages} {235} (\bibinfo {year}
  {2008})%
  \bibAnnoteFile{NoStop}{Luding08gm}%
\bibitem{tomas2004fundamentals}%
  \BibitemOpen
  \bibfield{author}{%
  \bibinfo {author} {\bibfnamefont{J.}~\bibnamefont{Tomas}},\ }%
  \bibfield{journal}{%
  \bibinfo {journal} {Granular Matter}\ }%
  \textbf{\bibinfo {volume} {6}},\ \bibinfo {pages} {75} (\bibinfo {year}
  {2004})%
  \bibAnnoteFile{NoStop}{tomas2004fundamentals}%
\bibitem{thornton1998theoretical}%
  \BibitemOpen
  \bibfield{author}{%
  \bibinfo {author} {\bibfnamefont{C.}~\bibnamefont{Thornton}}\ and\ \bibinfo
  {author} {\bibfnamefont{Z.}~\bibnamefont{Ning}},\ }%
  \bibfield{journal}{%
  \bibinfo {journal} {Powder Technology}\ }%
  \textbf{\bibinfo {volume} {12}} (\bibinfo {year} {1998})%
  \bibAnnoteFile{NoStop}{thornton1998theoretical}%
\bibitem{thakur2013experimental}%
  \BibitemOpen
  \bibfield{author}{%
  \bibinfo {author} {\bibfnamefont{S.~C.}\ \bibnamefont{Thakur}}, \bibinfo
  {author} {\bibfnamefont{H.}~\bibnamefont{Ahmadian}}, \bibinfo {author}
  {\bibfnamefont{J.}~\bibnamefont{Sun}},\ and\ \bibinfo {author}
  {\bibfnamefont{J.~Y.}\ \bibnamefont{Ooi}},\ }%
  \bibfield{journal}{%
  \bibinfo {journal} {Particuology}}%
   (\bibinfo {year} {2013})%
  \bibAnnoteFile{NoStop}{thakur2013experimental}%
\bibitem{pasha2014linear}%
  \BibitemOpen
  \bibfield{author}{%
  \bibinfo {author} {\bibfnamefont{M.}~\bibnamefont{Pasha}}, \bibinfo {author}
  {\bibfnamefont{S.}~\bibnamefont{Dogbe}}, \bibinfo {author}
  {\bibfnamefont{C.}~\bibnamefont{Hare}}, \bibinfo {author}
  {\bibfnamefont{A.}~\bibnamefont{Hassanpour}},\ and\ \bibinfo {author}
  {\bibfnamefont{M.}~\bibnamefont{Ghadiri}},\ }%
  \bibfield{journal}{%
  \bibinfo {journal} {Granular Matter}\ }%
  \textbf{\bibinfo {volume} {16}},\ \bibinfo {pages} {151} (\bibinfo {year}
  {2014})%
  \bibAnnoteFile{NoStop}{pasha2014linear}%
\bibitem{asingh14}%
  \BibitemOpen
  \bibfield{author}{%
  \bibinfo {author} {\bibfnamefont{A.}~\bibnamefont{Singh}}, \bibinfo {author}
  {\bibfnamefont{V.}~\bibnamefont{Magnanimo}},\ and\ \bibinfo {author}
  {\bibfnamefont{S.}~\bibnamefont{Luding}},\ }%
  \bibfield{journal}{%
  \bibinfo {journal} {Powder Technology}\ }%
  \textbf{\bibinfo {volume} {under review}} (\bibinfo {year} {2014})%
  \bibAnnoteFile{NoStop}{asingh14}%
\bibitem{fenistein04}%
  \BibitemOpen
  \bibfield{author}{%
  \bibinfo {author} {\bibfnamefont{D.}~\bibnamefont{Fenistein}}, \bibinfo
  {author} {\bibfnamefont{J.~W.}\ \bibnamefont{van~de Meent}},\ and\ \bibinfo
  {author} {\bibfnamefont{M.}~\bibnamefont{van Hecke}},\ }%
  \bibfield{journal}{%
  \bibinfo {journal} {Phys. Rev. Lett.}\ }%
  \textbf{\bibinfo {volume} {92}},\ \bibinfo {pages} {094301} (\bibinfo {year}
  {2004})%
  \bibAnnoteFile{NoStop}{fenistein04}%
\bibitem{luding08p}%
  \BibitemOpen
  \bibfield{author}{%
  \bibinfo {author} {\bibfnamefont{S.}~\bibnamefont{Luding}},\ }%
  \bibfield{journal}{%
  \bibinfo {journal} {Particulate Science and Technology}\ }%
  \textbf{\bibinfo {volume} {26}},\ \bibinfo {pages} {33} (\bibinfo {year}
  {2008})%
  \bibAnnoteFile{NoStop}{luding08p}%
\bibitem{luding08crs}%
  \BibitemOpen
  \bibfield{author}{%
  \bibinfo {author} {\bibfnamefont{S.}~\bibnamefont{Luding}},\ }%
  \bibfield{journal}{%
  \bibinfo {journal} {Particuology}\ }%
  \textbf{\bibinfo {volume} {6}},\ \bibinfo {pages} {501} (\bibinfo {year}
  {2008})%
  \bibAnnoteFile{NoStop}{luding08crs}%
\bibitem{dijksman10}%
  \BibitemOpen
  \bibfield{author}{%
  \bibinfo {author} {\bibfnamefont{J.~A.}\ \bibnamefont{Dijksman}}\ and\
  \bibinfo {author} {\bibfnamefont{M.}~\bibnamefont{van Hecke}},\ }%
  \bibfield{journal}{%
  \bibinfo {journal} {Soft Matter}\ }%
  \textbf{\bibinfo {volume} {6}},\ \bibinfo {pages} {2901} (\bibinfo {year}
  {2010})%
  \bibAnnoteFile{NoStop}{dijksman10}%
\bibitem{Unger04shear}%
  \BibitemOpen
  \bibfield{author}{%
  \bibinfo {author} {\bibfnamefont{T.}~\bibnamefont{Unger}}, \bibinfo {author}
  {\bibfnamefont{J.}~\bibnamefont{T\"or\"ok}}, \bibinfo {author}
  {\bibfnamefont{J.}~\bibnamefont{Kert\'esz}},\ and\ \bibinfo {author}
  {\bibfnamefont{D.~E.}\ \bibnamefont{Wolf}},\ }%
  \bibfield{journal}{%
  \bibinfo {journal} {Phys. Rev. Lett.}\ }%
  \textbf{\bibinfo {volume} {92}},\ \bibinfo {pages} {214301} (\bibinfo {year}
  {2004})%
  \bibAnnoteFile{NoStop}{Unger04shear}%
\bibitem{Fenistein06}%
  \BibitemOpen
  \bibfield{author}{%
  \bibinfo {author} {\bibfnamefont{D.}~\bibnamefont{Fenistein}}, \bibinfo
  {author} {\bibfnamefont{J.-W.}\ \bibnamefont{van~de Meent}},\ and\ \bibinfo
  {author} {\bibfnamefont{M.}~\bibnamefont{van Hecke}},\ }%
  \bibfield{journal}{%
  \bibinfo {journal} {Phys. Rev. Lett.}\ }%
  \textbf{\bibinfo {volume} {96}},\ \bibinfo {pages} {118001} (\bibinfo {year}
  {2006})%
  \bibAnnoteFile{NoStop}{Fenistein06}%
\bibitem{ries07}%
  \BibitemOpen
  \bibfield{author}{%
  \bibinfo {author} {\bibfnamefont{A.}~\bibnamefont{Ries}}, \bibinfo {author}
  {\bibfnamefont{D.~E.}\ \bibnamefont{Wolf}},\ and\ \bibinfo {author}
  {\bibfnamefont{T.}~\bibnamefont{Unger}},\ }%
  \bibfield{journal}{%
  \bibinfo {journal} {Phys. Rev. E}\ }%
  \textbf{\bibinfo {volume} {76}},\ \bibinfo {pages} {051301} (\bibinfo {year}
  {2007})%
  \bibAnnoteFile{NoStop}{ries07}%
\bibitem{da2005rheophysics}%
  \BibitemOpen
  \bibfield{author}{%
  \bibinfo {author} {\bibfnamefont{F.}~\bibnamefont{da~Cruz}}, \bibinfo
  {author} {\bibfnamefont{S.}~\bibnamefont{Emam}}, \bibinfo {author}
  {\bibfnamefont{M.}~\bibnamefont{Prochnow}}, \bibinfo {author}
  {\bibfnamefont{J.-N.}\ \bibnamefont{Roux}},\ and\ \bibinfo {author}
  {\bibfnamefont{F.}~\bibnamefont{Chevoir}},\ }%
  \bibfield{journal}{%
  \bibinfo {journal} {Phys. Rev. E}\ }%
  \textbf{\bibinfo {volume} {72}},\ \bibinfo {pages} {021309} (\bibinfo {year}
  {2005})%
  \bibAnnoteFile{NoStop}{da2005rheophysics}%
\bibitem{Ovarlez08}%
  \BibitemOpen
  \bibfield{author}{%
  \bibinfo {author} {\bibfnamefont{G.}~\bibnamefont{Ovarlez}}, \bibinfo
  {author} {\bibfnamefont{S.}~\bibnamefont{Rodts}}, \bibinfo {author}
  {\bibfnamefont{A.}~\bibnamefont{Ragouilliaux}}, \bibinfo {author}
  {\bibfnamefont{P.}~\bibnamefont{Coussot}}, \bibinfo {author}
  {\bibfnamefont{J.}~\bibnamefont{Goyon}},\ and\ \bibinfo {author}
  {\bibfnamefont{A.}~\bibnamefont{Colin}},\ }%
  \bibfield{journal}{%
  \bibinfo {journal} {Phys. Rev. E}\ }%
  \textbf{\bibinfo {volume} {78}},\ \bibinfo {pages} {036307} (\bibinfo {year}
  {2008})%
  \bibAnnoteFile{NoStop}{Ovarlez08}%
\bibitem{ABYu13}%
  \BibitemOpen
  \bibfield{author}{%
  \bibinfo {author} {\bibfnamefont{X.}~\bibnamefont{Wang}}, \bibinfo {author}
  {\bibfnamefont{H.~P.}\ \bibnamefont{Zhu}}, \bibinfo {author}
  {\bibfnamefont{S.}~\bibnamefont{Luding}},\ and\ \bibinfo {author}
  {\bibfnamefont{A.~B.}\ \bibnamefont{Yu}},\ }%
  \bibfield{journal}{%
  \bibinfo {journal} {Phys. Rev. E}\ }%
  \textbf{\bibinfo {volume} {88}},\ \bibinfo {pages} {032203} (\bibinfo {year}
  {2013})%
  \bibAnnoteFile{NoStop}{ABYu13}%
\bibitem{Shebani11}%
  \BibitemOpen
  \bibfield{author}{%
  \bibinfo {author} {\bibfnamefont{M.~R.}\ \bibnamefont{Shaebani}}, \bibinfo
  {author} {\bibfnamefont{M.}~\bibnamefont{Madadi}}, \bibinfo {author}
  {\bibfnamefont{S.}~\bibnamefont{Luding}},\ and\ \bibinfo {author}
  {\bibfnamefont{D.~E.}\ \bibnamefont{Wolf}},\ }%
  \bibfield{journal}{%
  \bibinfo {journal} {Phys. Rev. E}\ }%
  \textbf{\bibinfo {volume} {85}},\ \bibinfo {pages} {011301} (\bibinfo {year}
  {2012})%
  \bibAnnoteFile{NoStop}{Shebani11}%
\bibitem{Luding2005}%
  \BibitemOpen
  \bibfield{author}{%
  \bibinfo {author} {\bibfnamefont{S.}~\bibnamefont{Luding}}, \bibinfo {author}
  {\bibfnamefont{K.}~\bibnamefont{Manetsberger}},\ and\ \bibinfo {author}
  {\bibfnamefont{J.}~\bibnamefont{M{\"u}llers}},\ }%
  \bibfield{journal}{%
  \bibinfo {journal} {Jour. Mech. and Phys. Solids}\ }%
  \textbf{\bibinfo {volume} {53}},\ \bibinfo {pages} {455} (\bibinfo {year}
  {2005})%
  \bibAnnoteFile{NoStop}{Luding2005}%
\bibitem{Hecke_tail05}%
  \BibitemOpen
  \bibfield{author}{%
  \bibinfo {author} {\bibfnamefont{A.~R.~T.}\ \bibnamefont{van Eerd}}, \bibinfo
  {author} {\bibfnamefont{W.~G.}\ \bibnamefont{Ellenbroek}}, \bibinfo {author}
  {\bibfnamefont{M.}~\bibnamefont{van Hecke}}, \bibinfo {author}
  {\bibfnamefont{J.~H.}\ \bibnamefont{Snoeijer}},\ and\ \bibinfo {author}
  {\bibfnamefont{T.~J.~H.}\ \bibnamefont{Vlugt}},\ }%
  \bibfield{journal}{%
  \bibinfo {journal} {Phys. Rev. E}\ }%
  \textbf{\bibinfo {volume} {75}},\ \bibinfo {pages} {060302} (\bibinfo {year}
  {2007})%
  \bibAnnoteFile{NoStop}{Hecke_tail05}%
\bibitem{radjai1996force}%
  \BibitemOpen
  \bibfield{author}{%
  \bibinfo {author} {\bibfnamefont{F.}~\bibnamefont{Radjai}}, \bibinfo {author}
  {\bibfnamefont{M.}~\bibnamefont{Jean}}, \bibinfo {author}
  {\bibfnamefont{J.-J.}\ \bibnamefont{Moreau}},\ and\ \bibinfo {author}
  {\bibfnamefont{S.}~\bibnamefont{Roux}},\ }%
  \bibfield{journal}{%
  \bibinfo {journal} {Phys.\ Rev.\ Lett.}\ }%
  \textbf{\bibinfo {volume} {77}},\ \bibinfo {pages} {274} (\bibinfo {year}
  {1996})%
  \bibAnnoteFile{NoStop}{radjai1996force}%
\bibitem{makse2000packing}%
  \BibitemOpen
  \bibfield{author}{%
  \bibinfo {author} {\bibfnamefont{H.~A.}\ \bibnamefont{Makse}}, \bibinfo
  {author} {\bibfnamefont{D.~L.}\ \bibnamefont{Johnson}},\ and\ \bibinfo
  {author} {\bibfnamefont{L.~M.}\ \bibnamefont{Schwartz}},\ }%
  \bibfield{journal}{%
  \bibinfo {journal} {Physical review letters}\ }%
  \textbf{\bibinfo {volume} {84}},\ \bibinfo {pages} {4160} (\bibinfo {year}
  {2000})%
  \bibAnnoteFile{NoStop}{makse2000packing}%
\bibitem{zhang2005jamming}%
  \BibitemOpen
  \bibfield{author}{%
  \bibinfo {author} {\bibfnamefont{H.}~\bibnamefont{Zhang}}\ and\ \bibinfo
  {author} {\bibfnamefont{H.}~\bibnamefont{Makse}},\ }%
  \bibfield{journal}{%
  \bibinfo {journal} {Physical Review E}\ }%
  \textbf{\bibinfo {volume} {72}},\ \bibinfo {pages} {011301} (\bibinfo {year}
  {2005})%
  \bibAnnoteFile{NoStop}{zhang2005jamming}%
\bibitem{Brian05force}%
  \BibitemOpen
  \bibfield{author}{%
  \bibinfo {author} {\bibfnamefont{B.~P.}\ \bibnamefont{Tighe}}, \bibinfo
  {author} {\bibfnamefont{J.~E.~S.}\ \bibnamefont{Socolar}}, \bibinfo {author}
  {\bibfnamefont{D.~G.}\ \bibnamefont{Schaeffer}}, \bibinfo {author}
  {\bibfnamefont{W.~G.}\ \bibnamefont{Mitchener}},\ and\ \bibinfo {author}
  {\bibfnamefont{M.~L.}\ \bibnamefont{Huber}},\ }%
  \bibfield{journal}{%
  \bibinfo {journal} {Phys. Rev. E}\ }%
  \textbf{\bibinfo {volume} {72}},\ \bibinfo {pages} {031306} (\bibinfo {year}
  {2005})%
  \bibAnnoteFile{NoStop}{Brian05force}%
\bibitem{tahmasebpoora13}%
  \BibitemOpen
  \bibfield{author}{%
  \bibinfo {author} {\bibfnamefont{M.}~\bibnamefont{Tahmasebpoora}}, \bibinfo
  {author} {\bibfnamefont{L.}~\bibnamefont{de~Martin}}, \bibinfo {author}
  {\bibfnamefont{M.}~\bibnamefont{Talebi}}, \bibinfo {author}
  {\bibfnamefont{N.}~\bibnamefont{Mostoufia}},\ and\ \bibinfo {author}
  {\bibfnamefont{J.~R.}\ \bibnamefont{van Ommen}},\ }%
  \bibfield{journal}{%
  \bibinfo {journal} {Phys. Chem. Chem. Phys.}\ }%
  \textbf{\bibinfo {volume} {15}},\ \bibinfo {pages} {5788} (\bibinfo {year}
  {2013})%
  \bibAnnoteFile{NoStop}{tahmasebpoora13}%
\bibitem{fuchs2014}%
  \BibitemOpen
  \bibfield{author}{%
  \bibinfo {author} {\bibfnamefont{R.}~\bibnamefont{Fuchs}}, \bibinfo {author}
  {\bibfnamefont{T.}~\bibnamefont{Weinhart}}, \bibinfo {author}
  {\bibfnamefont{J.}~\bibnamefont{Meyer}}, \bibinfo {author}
  {\bibfnamefont{T.}~\bibnamefont{Staedler}}, \bibinfo {author}
  {\bibfnamefont{X.}~\bibnamefont{Jiang}},\ and\ \bibinfo {author}
  {\bibfnamefont{S.}~\bibnamefont{Luding}},\ }%
  \bibfield{journal}{%
  \bibinfo {journal} {Granular Matter}}%
   (\bibinfo {year} {2014})%
  \bibAnnoteFile{NoStop}{fuchs2014}%
\end{thebibliography}%
\end{document}